# Quantification of spin-charge interconversion in highly resistive sputtered $Bi_xSe_{1-x}$ with non-local spin valves

Isabel C. Arango[1,5*], Won Young Choi[1,5], Van Tuong Pham[2], Inge Groen[1], Diogo C. Vaz[1], Punyashloka Debashis[4], Hai Li[4], Mahendra DC[4], Kaan Oguz[4], Andrey Chuvilin[1,3], Luis E. Hueso[1,3], Ian A. Young[4] and Fèlix Casanova[1,3*]

[1]CIC nanoGUNE BRTA, 20018 Donostia-San Sebastián, Basque Country, Spain
[2]IMEC, Kapeldreef 75, 3001 Leuven, Belgium
[3]IKERBASQUE, Basque Foundation for Science, 48009 Bilbao, Basque Country, Spain
[4]Components Research, Intel Corp., Hillsboro, Oregon 97124, United States
[5]These authors contributed equally: Isabel C. Arango, Won Young Choi.
*e-mail: i.arango@nanogune.eu, f.casanova@nanogune.eu

## Abstract

The development of spin-orbitronic devices, such as magneto-electric spin-orbit logic devices, calls for materials with a high resistivity and a high spin-charge interconversion efficiency. One of the most promising candidates in this regard is sputtered $Bi_xSe_{1-x}$. Although there are several techniques to quantify spin-charge interconversion, to date reported values for sputtered $Bi_xSe_{1-x}$ have often been overestimated due to spurious effects related to local currents combined with a lack of understanding of the effect of the interfaces and the use of approximations for unknown parameters, such as the spin diffusion length. In the present study, non-local spin valves are used to inject pure spin currents into $Bi_xSe_{1-x}$, allowing us to directly obtain its spin diffusion length as well as its spin Hall angle, from 10 K up to 300 K. These values, which are more accurate than those previously reported in sputtered $Bi_xSe_{1-x}$, evidence that the efficiency of this material is not exceptional. Indeed, the figure of merit for spin-charge interconversion, given by the product of these two parameters, is slightly under 1 nm. Our work demonstrates the importance of considering all material parameters and interfaces when quantifying the spin transport properties of materials with strong spin-orbit coupling.

## I. INTRODUCTION

Moore's 1965 prediction on the downscaling of transistor prediction [1,2] has held strong for a remarkable amount of time. However, the CMOS technology on which it has relied thus far is now reaching its scaling limits, sparking an intense effort [3] to find alternative approaches with new functionalities that can be integrated in the next generation of electronic devices. One of these approaches, known as spintronics, makes use of the electron's spin degree of freedom in non-volatile memories [4,5] and logic devices [6–8]. Spintronics relies on materials with strong spin-orbit coupling (SOC), which allow spin-charge interconversion (SCI) via the spin Hall effect (SHE) [9] or the Edelstein effect [10]. A recent proposal in the field introduced a new device concept known as magneto-electric spin-orbit (MESO) [8,11] for logic operations based on one device with two different nodes cascading multiple devices. The input node is used to write a magnetic element with voltage using magnetoelectric effects and the output node to read the magnetic state of the element with spin-to-charge conversion [12]. One of the requirements for MESO is that the readout voltage should be above the coercive voltage of the magnetoelectric material (~100 mV) to drive the next element in a logic operation, a value which could be reached by using materials with high SCI efficiencies and high resistivities [13].

In this regard, one promising candidate for the readout node of the MESO device is $Bi_2Se_3$, which has been reported to have both high SCI efficiency and high resistivity [14,15]. $Bi_2Se_3$ is well-known as a topological insulator [16]. This class of materials shows spin-momentum locking at the topologically protected surface states, a feature that allows an efficient Edelstein effect (characterized by the inverse Edelstein length, $\lambda_{IEE}$). Exploitation of these surface states typically requires an epitaxial structure and low temperature to minimize bulk conduction [17,18]. Recently, however, some works [19–23] reported large SCI even at room temperature in polycrystalline $Bi_xSe_{1-x}$ (BiSe) grown by sputtering, a simple technique compatible with the industrial processes. Although the Edelstein effect is the source of SCI in ideal topological insulators, many works use the spin Hall angle ($\theta_{SH}$) to quantify the SCI efficiency in this class of materials [24]. In this case, just like for materials exhibiting SHE, the spin diffusion length ($\lambda_s$) is an essential parameter for a proper quantification of the SCI efficiency. Indeed, for many applications including MESO, the relevant figure of merit is the $\theta_{SH}\lambda_s$ product [13], which is equivalent to $\lambda_{IEE}$ [25]. However, $\lambda_s$ for sputtered BiSe is usually taken from few reports describing epitaxially grown $Bi_2Se_3$ [26–29], which not only has a different crystal structure, but also a different composition. This dissimilarity invariably leads to inaccuracies in the subsequent quantification of the SCI efficiency of BiSe. In addition, most SCI quantification techniques require the SCI material to be in direct contact with a ferromagnetic or transition metal, but recent studies on $Bi_2Se_3$ have reported interdiffusion by solid-state reaction when it is in contact with metals [30–34]. Thus, a new layer forms at the interface through which the spins are injected or pumped, making an accurate quantification of the spin properties ($\lambda_s$ and $\theta_{SH}$) of this material difficult [31,34].

In this article, we characterize sputtered BiSe through the spin absorption technique using lateral spin valves (LSVs) and two separate measurement configurations [35–38]. This non-local method allows us to independently quantify the spin diffusion length ($\lambda_s^{BiSe}$) and the spin Hall angle ($\theta_{SH}^{BiSe}$) of BiSe. The use of a non-local measurement avoids spurious effects related to local currents, such as Oersted fields in spin-orbit torque techniques or fringe-field-induced voltages in three-terminal potentiometric techniques [39]. Furthermore, in order to reduce interdiffusion, we grow the metals in contact with the BiSe wire by e-beam evaporation, a gentler deposition technique than sputtering. A much better quality of the device interface is confirmed by transmission electron microscopy (TEM) and elemental analysis characterization. This information allows us to model our devices and perform a 3D Finite Element Method (3D FEM) analysis to extract the spin transport parameters at different temperatures. The SCI efficiency, characterized by the $\theta_{SH}^{BiSe}\lambda_s^{BiSe}$ product, is found to be up to 0.92 nm at 100 K, and 0.63 nm at room temperature. Our work highlights the importance of considering all the details of BiSe and its interfaces for a proper quantification of the spin transport properties of this material.

## II. EXPERIMENTAL DETAILS

### a. Device fabrication

All LSV devices were fabricated on $Si/SiO_2$ (150 nm) substrates (see Fig. 1a). Three electron-beam lithography (eBL) steps are needed for the complete fabrication of the LSVs. The first step is used to define the ferromagnetic wires: we spin-coated the substrates using ZEP (methyl styrene and chloromethyl acrylate copolymer) as a positive resist, patterned it by eBL, deposited 30 nm of Py ($Ni_{81}Fe_{19}$) by e-beam evaporation (base pressure of 2×10$^{-9}$ Torr, rate of 0.6 Å/s), and performed the lift-off process. To remove possible sidewalls on the wires after the lift-off, we Ar-ion milled the sample at an angle of 10º with respect to the substrate plane and an acceleration voltage of 50 V. The second step defines the BiSe wires: we spin-coated a double layer of PMMA (polymethyl methacrylate), patterned it by eBL, deposited 10 nm of BiSe by sputtering at room temperature, using a target of stoichiometric $Bi_2Se_3$ (99.999% purity) in a UHV seven-target AJA sputtering system with a base pressure of 3×10$^{-8}$ Torr. $Bi_2Se_3$ was sputtered at 35 W RF power and 3 mTorr Ar pressure, yielding a deposition rate of 0.09 Å/s. Subsequently, the wires were capped *in situ* with 2 nm of Pt (80 W DC at 3 mTorr Ar pressure) and lift-off was performed. The third step defines the Cu spin transport channel: we used a double layer of PMMA, patterned it with eBL, and then used Ar-ion milling to remove the Pt capping and clean the surfaces of the Py wires. We then transferred the sample to the UHV evaporation system to grow 2 nm of Ti by e-beam evaporation (at a rate of 0.2 Å/s), followed by 100 nm of Cu *in situ* by thermal evaporation at a rate of 1.5 Å/s. The Ti layer is added to help the Cu grow on top of BiSe (see Supplemental Material S1 [40]) and acts as an interface between the Cu channel and the Py and BiSe electrodes. Lift-off was then performed. Finally, the entire sample was capped by sputtering 5 nm of $SiO_2$ (200 W RF at 3 mTorr of Ar).

### b. **Transport measurements**

Transport measurements were performed in a Quantum Design Physical Properties Measurement System (PPMS), using the "dc reversal" technique with a Keithley 2182 nanovoltmeter and a 6221 current source. Thermoelectric effects arising from Joule heating are removed with the use of the "dc reversal" technique [41].

### c. **TEM characterization**

Cross-sectional samples for analysis by scanning transmission electron microscopy with energy dispersive X-ray spectroscopy (STEM-EDX) were prepared from tested devices by the standard focused ion beam (FIB) lamella preparation method: the surface of the deposited samples was first protected by ion beam Pt deposition, the lamellas were cut and lifted onto a Cu 3-post half-grid. Cross sections were studied on a Titan 60-300 TEM (FEI, Netherlands) at 300 kV in STEM mode. EDX spectral images were obtained using an EDAX RTEM spectrometer. Element distribution maps were obtained by multiple linear least squares (MLLS) fitting of experimental spectra using simulated spectral components.

## III. RESULTS and DISCUSSION

### a. Lateral spin valves

In a reference LSV without the BiSe wire, a charge current ($I_C$) is injected from one of the ferromagnetic Py electrodes into the non-magnetic Cu channel, creating a spin accumulation at the interface. These spins diffuse as a pure spin current ($I_s$) through the Cu channel with a characteristic diffusion length ($\lambda_s^{Cu}$) and is detected by the second Py electrode as a non-local voltage ($V_{NL}$). The non-local resistance, $R_{NL}$, is defined as the $V_{NL}$ normalized to $I_C$. An external magnetic field is applied along the easy axis of the ferromagnet (± *y*-direction) to control the reversal of the magnetization of the two Py electrodes. The value of $R_{NL}$ changes sign when the magnetization configuration of the two Py electrodes switches from parallel ($R_{NL}^P$) to antiparallel ($R_{NL}^{AP}$). The difference between these two configurations ($\Delta R_{NL}^{Ref} = R_{NL}^{P} - R_{NL}^{AP}$) allows us to obtain the spin signal by removing any baseline arising from non-spin related effects.

In a similar LSV device, we place a BiSe wire between the two Py electrodes (Fig. 1a). Part of the spin current diffusing along the Cu channel will be absorbed in the BiSe wire and, thus, the spin signal picked up by the Py detector, $\Delta R_{NL}^{Abs}$, will be smaller than $\Delta R_{NL}^{Ref}$ (see Fig. 1b). The Ti/Cu cross is deposited on top of the BiSe wire to improve the electrical contact due to the high resistivity of this material [31], and to help us perform spin-to-charge conversion measurements on the same device (see below). Figure 1c plots the values of $\Delta R_{NL}^{Ref}$ and $\Delta R_{NL}^{Abs}$ at different temperatures ($T$) between 10 and 300 K. The decrease of the spin signals with increasing $T$ is expected because $\lambda_s^{Cu}$ decreases with temperature (ref. [42] and Supplemental Material S3 [40]) and less spin current reaches the ferromagnetic detector.

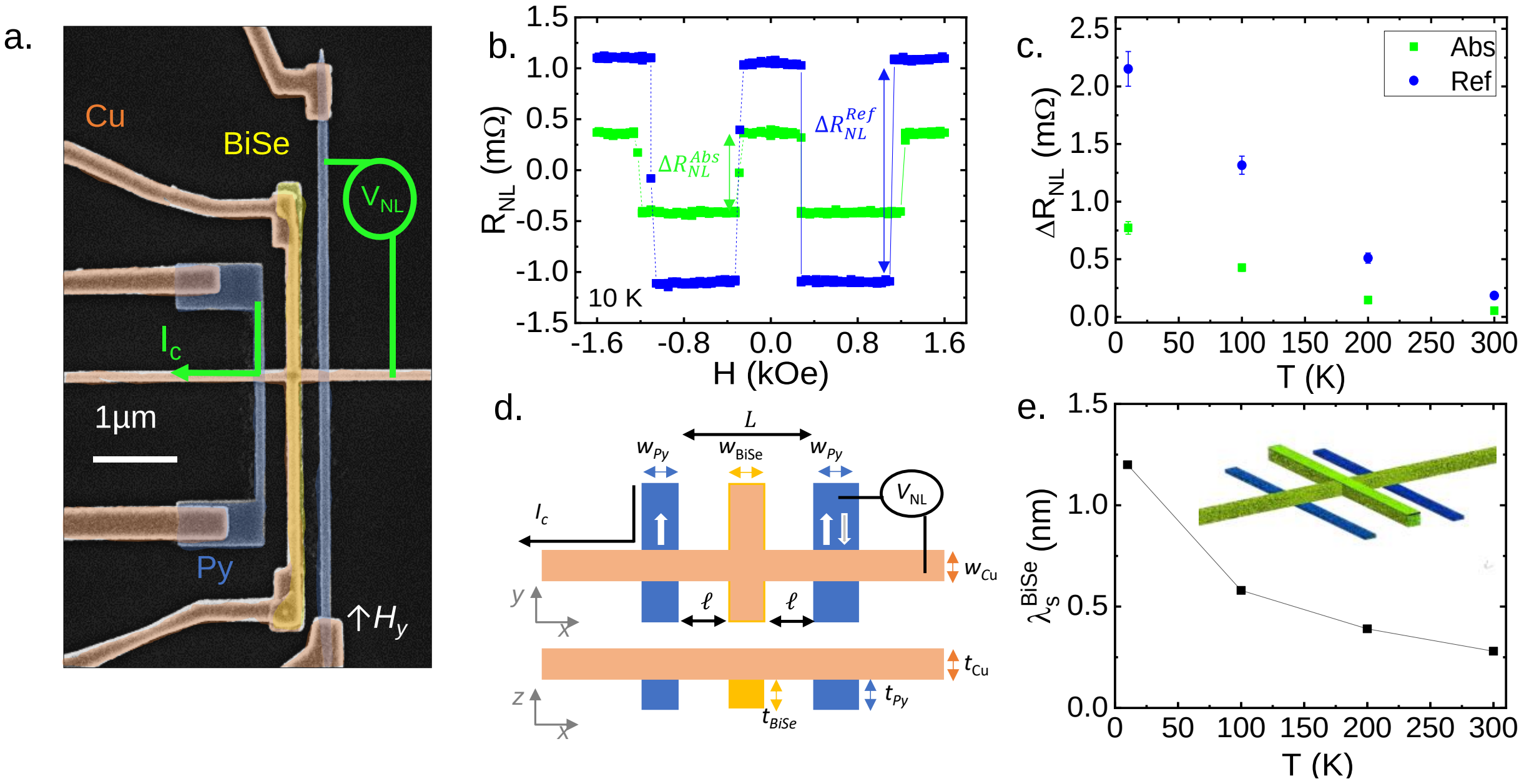

**Fig. 1 | a.** Top-view SEM image of the LSV device with the ferromagnetic Py electrodes (blue), the BiSe wire in between (yellow), and the Cu spin transport channel (orange). The electrical configuration for the spin absorption measurement is shown in green. **b.** Non-local resistance $R_{NL}$ as a function of the external magnetic field for the reference LSV (blue curve) and the LSV with the BiSe wire (green curve) at 10 K. The corresponding spin signals are indicated by arrows. **c.** Spin signals as a function of temperature for the reference LSV and the LSV with the BiSe wire. **d.** Schematic representation of the LSV with the BiSe wire showing the geometrical parameters in the top view (upper part) and cross-sectional view (bottom part). The ferromagnetic electrodes are separated by a distance $L = 650$ nm. **e.** Spin diffusion length of BiSe, $\lambda_s^{BiSe}$, as a function of the temperature, extracted from the data in panel c through a 3D FEM analysis. In the 3D FEM simulation, the resistivity of the Ti layer between Cu and BiSe is set to be 50 μΩcm. Inset: Geometry and mesh of the 3D FEM model.

To extract the spin diffusion length of BiSe ($\lambda_s^{BiSe}$) from the spin absorption measurement, we performed a 3D FEM simulation using: i) the experimental resistivities for all materials (Supplemental Material S2 [40]); ii) the interface spin polarization ($\alpha_I$) of the Py/Ti and $\lambda_s^{Cu}$ of the Cu channel, obtained from reference LSVs with different electrode distances ($L$) by fitting the spin signals $\Delta R_{NL}$ vs $L$ using the 1D spin diffusion model (see Supplemental Material S3 [40]); iii) the contact resistance of the Py/Ti/Cu interface extracted from an interface resistance measurement with four probe configuration (Supplemental Material

S4 [40]); iv) the measured spin signal after absorption ($\Delta R_{NL}^{Abs}$). Besides $\lambda_s^{BiSe}$, the only unknown parameter in the 3D FEM simulation is the resistivity of the Ti layer ($\rho_{Ti}$) between Cu and BiSe. Unfortunately, due to our device geometry, with the Ti layer sandwiched between the Cu channel and the BiSe wire, it is not possible to extract directly $\rho_{Ti}$. Therefore, we estimate this resistivity in a separate experiment described in Supplemental Material S5 [40], from which we obtain the value 50 μΩcm. As described in Supplemental Material S6 [40], we extracted $\lambda_s^{BiSe}$ at different temperatures, as shown in Fig. 1e. The extracted value is in all cases of the order or smaller than 1 nm, reaching 0.28 nm at room temperature. This value is significantly smaller than the values of 1.6 to 6.2 nm previously reported for epitaxially grown $Bi_2Se_3$ [26,27].

### b. Spin-charge interconversion

In the same device used to performed the spin absorption measurement, we measure the inverse spin Hall effect (ISHE) using a different electrical configuration (see sketch in Fig. 2a). This time, we inject a charge current ($I_C$) from one of the Py electrodes into the Cu channel while applying an in-plane magnetic field along the hard axis of Py ($x$-direction). An $x$-polarized spin current is created and reaches the BiSe wire, where it is partially absorbed in the $z$-direction and converted into a charge current ($I_{ISHE}$) along the $y$-direction (Fig. 2c). This charge current is detected as a voltage ($V_{ISHE}$) along the BiSe wire (shunted by Cu) under open-circuit conditions. The ISHE resistance ($R_{ISHE} = V_{ISHE}/I_C$) is measured by sweeping the external magnetic field along the $x$-direction. By reversing the field, the opposite $R_{ISHE}$ is obtained (see Fig. 2d), because the Py magnetization changes direction and, thus, so does the spin polarization of the spin current. The difference between the two $R_{ISHE}$ values at saturation is denoted as the ISHE signal ($2\Delta R_{ISHE}$) and allows removing any background signal. Indeed, the combination of Seebeck and Peltier effects can give rise to a baseline in the non-local signal because they are linear with the applied current, as explained in Ref. [43]. However, this contribution is removed by taking the difference between the two values at saturation. Since the material we study is not magnetic, a spurious contribution due to the combination of Peltier effect and anomalous Nernst effect as the one observed in the Weyl ferromagnet $Co_2MnGa$ [44] is not present in our case.

As shown in Fig. 2b, it is possible to obtain the direct SHE by swapping the current and voltage probes (i.e., applying the charge current in the BiSe wire and measuring the output voltage between the Py electrode and the Cu channel). Both SHE and ISHE resistance curves have the same amplitude but opposite sign ($R_{ISHE}(H) = R_{SHE}(-H)$, see Fig. 2d), as expected from Onsager's reciprocity [45,46]. The (I)SHE signals decrease with increasing $T$, as shown in Fig. 2e. To extract the spin Hall angle ($\theta_{SH}^{BiSe}$) from the ISHE measurement, we performed a 3D FEM simulation using the same geometry (dimensions) and material parameters as before, plus the $\lambda_s^{BiSe}$ value obtained in the very same device (see Supplemental Material S7 [40]). Figure 2f shows the $\theta_{SH}^{BiSe}$ values extracted as a function of $T$ (from 0.69 at 10 K to 2.26 at 300 K). These values are more accurate than those previously reported in sputtered BiSe because of the knowledge of $\lambda_s^{BiSe}$. As a control experiment, we measured the reference device (without the BiSe) with the direct spin Hall configuration. As expected, no signal is observed (see Supplemental Material S8 [40]).

The product $\theta_{SH}^{BiSe}\lambda_s^{BiSe}$, which is the figure of merit for the efficiency of a MESO device, is shown in Fig. 2g as a function of $T$. $\theta_{SH}^{BiSe}\lambda_s^{BiSe}$ does not change significantly with $T$, having values between 0.63 and 0.92 nm, slightly higher than the prototypical heavy metals [13,35,47,48].

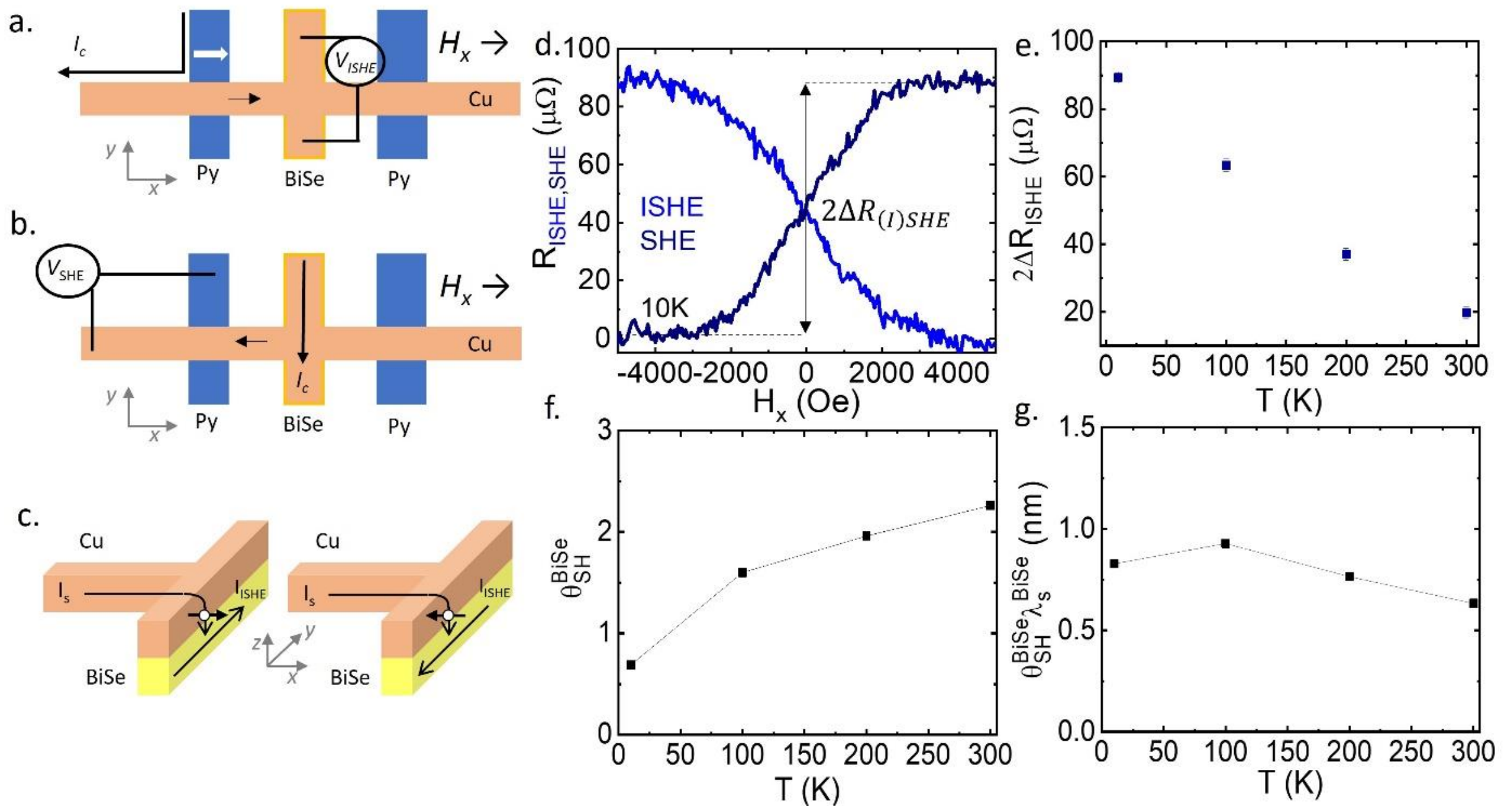

**Fig. 2 |** Schematic representation of the LSV with the middle wire showing **a.** spin-to-charge (ISHE) and **b.** charge-to-spin (SHE) conversion measurement configurations. **c.** After the spin injection from the Py (blue) to the Cu spin transport channel (orange) in panel a, the spin current flows (with a spin polarization in the $+x$-direction) along the Cu channel and is absorbed by the BiSe wire (yellow) along the $-z$-direction. Due to the ISHE, the spin current is converted into a charge current ($I_{ISHE}$) in the $+y$-direction (left sketch). If the magnetic field is reversed, the spin polarization of the spin current also reverses ($-x$-direction) and as does the converted $I_{ISHE}$ ($-y$-direction) (right sketch), which is detected as an open-circuit voltage. **d.** ISHE ($R_{ISHE}$) and SHE ($R_{SHE}$) resistances as function of the external magnetic field at 10 K. The corresponding (I)SHE signal is indicated by an arrow. **e.** ISHE signal ($2\Delta R_{ISHE}$) as a function of temperature. **f.** Spin Hall angle of BiSe ($\theta_{SH}^{BiSe}$) extracted from the data in panel e and a 3D FEM analysis, considering a Ti resistivity of 50 μΩcm. **g.** $\theta_{SH}^{BiSe}\lambda_s^{BiSe}$ product as a function of temperature.

### c. Transmission electron microscopy

After the magnetotransport characterization (spin absorption and spin Hall measurements), we characterized the device cross-sections by TEM/STEM imaging combined with EDX analysis with particular emphasis on materials interfaces. Interfaces play a key role in spintronic and spin-orbitronic devices, since they can enhance or reduce the efficiency of the spin current injection in SCI experiments [33,34,49–53]. Figure 3a shows a cross-sectional view of the LSV. The two Py electrodes (spin injector and detector) can be observed at the right and left of the image with the 10-nm-thick BiSe wire between them. They are covered by a homogeneous 2-nm-thick Ti layer followed by the 100-nm-thick Cu channel. The chemical distribution has been characterized by EDX. Figure 3b shows the different elemental maps for the elements of interest obtained by EDX in the region indicated by the orange rectangle in Fig. 3a (additional information in Supplemental Material S9 [40]). The elemental maps evidence that the 2-nm-thick Ti buffer layer is oxidized throughout the device. Figure 3d shows a higher resolution image of the area defined by the blue rectangle in Fig. 3c: the chemical distribution of Bi and Se within the BiSe wire evidences that the two elements are not homogeneously distributed and suggests diffusion has taken place inside the wire. A universal characterization of sputtered BiSe may become difficult due to the unavoidable fact that it is a highly reactive material in contact with other metallic materials. Figure 3e shows a high-resolution TEM image of the same wire (red rectangle in Fig. 3c), where the polycrystalline and granular structure of the BiSe layer can be observed. In some grains alternating $Bi_2Se_3$ quintuple layers and Bi bilayers are visible, in agreement with a previous report [31]. The layer of Ti on top of BiSe can be clearly distinguished and shows an amorphous morphology. Since, after Ar-ion milling, the Ti and then the Cu spin transport channel are deposited *ex situ* by e-beam evaporation, the interface does not show a detectable interdiffusion, in contrast to what is reported by contacting BiSe with transition metals by sputter deposition (all *in situ*) [31,34] and molecular-beam-epitaxy-grown $Bi_2Se_3$ with metallic contacts deposited by e-beam evaporation [30].

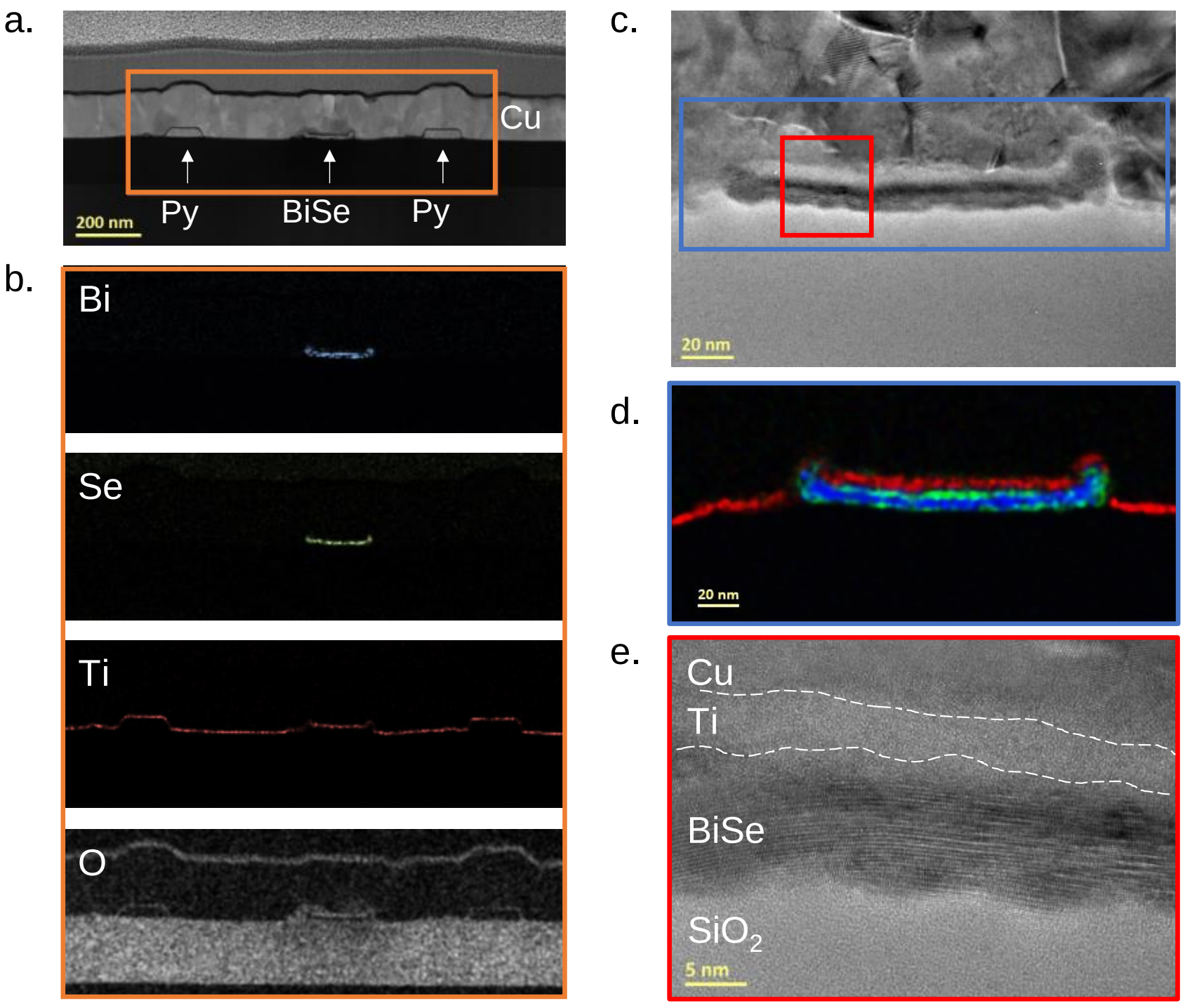

**Fig. 3 | a.** Cross-sectional TEM image of the LSV device. The orange rectangle indicates the area where EDX analysis was performed. **b.** EDX analysis of the cross-section of the LSV (orange rectangle in panel a), showing the elements of interest in each subpanel: Bi (blue), Se (green), Ti (red), and O (grey). **c.** Cross-sectional TEM image of the BiSe wire inside the LSV. **d.** Color-coded of elements image corresponding to the area marked with a blue rectangle in panel c. **e.** High resolution TEM image corresponding to the red rectangle area in panel c.

### d. $TiO_x$ interface layer

In the 3D FEM simulations described above and performed for both spin absorption and spin Hall measurements, we considered a metallic Ti layer with $\rho_{Ti}$=50 μΩcm, a value obtained from our control experiment (Supplemental Material S5 [40]). However, as pointed out in the previous section, our EDX analysis shows that the Ti layer in the LSV becomes oxidized. Therefore, to obtain more accurate values of $\lambda_S^{BiSe}$ and $\theta_{SH}^{BiSe}$, we need account for the presence of oxygen by increasing the resistivity of Ti. As mentioned, our device geometry does not allow extraction of the resistivity of the Ti layer in contact with the BiSe, however since the Ti layer also covers the Py electrodes, we were able to measure the interface resistance at that junction using the four probe configuration (Supplemental Material S4 [40]). Taking this value and calculating the resistivity for the 2-nm-thick oxidized Ti layer, we obtained $\rho_{Ti}$≈1000 μΩcm. The same material will grow differently on different materials and, therefore, we cannot directly assume that the resistivity of Ti on Py will be the same as that of Ti on BiSe, but we can take it as an upper limit. Repeating the 3D FEM simulation with $\rho_{Ti}$ values from 50 to 1000 μΩcm (see Supplemental Material S10 [40]) we extracted $\lambda_s^{BiSe}$ as a function of $\rho_{Ti}$, which is plotted in Fig. 4a. At 10 K, for example, $\lambda_s^{BiSe}$ varies between 1.1 and 2.7 nm. In order to rule out the possibility of a higher resistivities of the Ti layer, we also performed a simulation using $\rho_{Ti}$=1500 μΩcm (see Supplemental Material S10 [40]). In this case, $\lambda_s^{BiSe}$ tends to infinity, that is, fewer spins can reach the BiSe layer, rendering the properties of this second layer irrelevant in the 3D model. We also performed a simulation considering the BiSe resistivity measured in the vertical direction (across the thickness) at 300 K by Choi *et al.* ($\rho_{BiSe}$= 600 μΩcm) [31] (see Supplemental Material S10 [40]), a much lower value compared to our own measured value at room temperature ($\rho_{BiSe}$= 4100 μΩcm, see Supplemental Material S2 [40]). Comparison of the extracted spin diffusion length for the simulations using our measured values of $\rho_{BiSe}$ (see Fig. 4a, blue, black, red and light green curves) shows $\lambda_s^{BiSe}$ to be relatively small in all cases, and to decrease with increasing temperature for any $\rho_{Ti}$. However, comparing $\lambda_s^{BiSe}$ at room temperature (light green and dark cyan curves) and $\rho_{Ti}$ lower than 100 μΩcm, we find that the low BiSe resistivity value (~6 times smaller) yields a spin diffusion length more than three times larger than the one obtained using our higher value ($\rho_{BiSe}$= 4100 μΩcm).

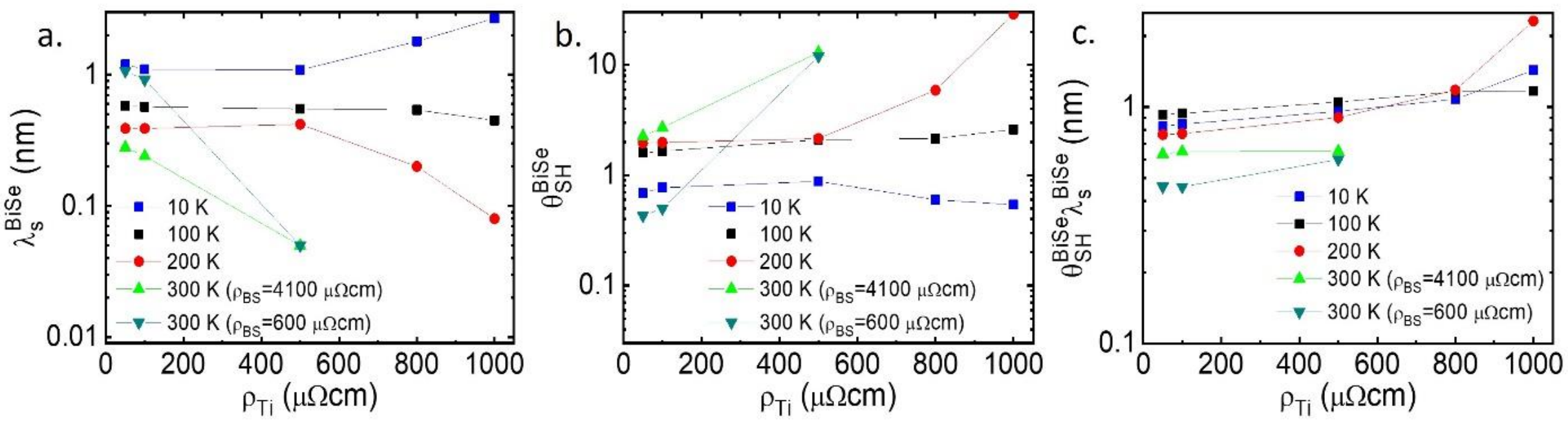

**Fig. 4 | a.** Spin diffusion length, $\lambda_s^{BiSe}$, **b.** spin Hall angle, $\theta_{SH}^{BiSe}$, and **c.** the $\theta_{SH}^{BiSe}\lambda_s^{BiSe}$ product extracted from the 3D FEM analysis as a function of the Ti resistivity ($\rho_{Ti}$) at different temperatures, from 10 K up to 300 K. Additionally, at 300 K, we also use a lower resistivity of BiSe reported in Ref. [31].

We additionally performed a 3D FEM simulation (see Supplemental Material S11 [40]) to extract the conversion efficiency ($\theta_{SH}^{BiSe}$), using the new values of $\lambda_s^{BiSe}$ for each value of $\rho_{Ti}$ from 50 to 1000 μΩcm. Our simulation results for $\theta_{SH}^{BiSe}$ as a function of $\rho_{Ti}$ are plotted in Fig. 4b. As an example, the value at 10 K varies between 0.54 and 0.88. We also performed the simulation considering $\rho_{BiSe}$= 600 μΩcm at 300 K (Supplemental Material S11 [40]). Finally, the product $\theta_{SH}^{BiSe}\lambda_s^{BiSe}$ as a function of $\rho_{Ti}$ is shown in Fig. 4c. Interestingly, this product does not present large variations with the resistivity of the interface layer, being fairly constant and generally lower than 1 nm. This indicates the robustness of this figure of merit in our analysis independently of the assumed resistivity of the interfacial Ti. Table 1 summarizes our results, taking the minimum and maximum values of these parameters for each temperature. Results from previous reports on sputtered BiSe are also included for comparison.

**Table 1 |** Summary of $\rho_{BiSe}$, $\lambda_s^{BiSe}$, $\theta_{SH}^{BiSe}$, $\lambda_{IEE}$ values obtained in this work and in previous reports on sputtered BiSe.

| | $t_{BiSe}\ (nm)$ | T (K) | $\rho_{BiSe}$ $(\mu\Omega cm)$ | $\lambda_s^{BiSe}\ (nm)$ | $\theta_{SH}^{BiSe}$ | $\lambda_{IEE}\ (nm)$ | Method |
|---|---|---|---|---|---|---|---|
| **$Bi_xSe_{1-x}$ (This work)** | 10 | 10 | 6200 | 1.09 - 2.70 | 0.54 - 0.88 | 0.82 - 1.46[a] | Non-local device (LSV) |
| | | 100 | 5900 | 0.45 - 0.58 | 1.60 - 2.61 | 0.92 - 1.17[a] | |
| | | 200 | 5100 | 0.08 - 0.39 | 1.96 - 28.90 | 0.76 - 2.31[a] | |
| | | 300 | 4100 | 0.05 - 0.28 | 2.26 – 13.01 | 0.63 - 0.65[a] | |
| | | 300 | 600[b] | 0.05 - 1.07 | 0.43 – 11.99 | 0.46 - 0.60[a] | |
| **$Bi_{50}Se_{50}$/Ti** [31] | 2-16 | 300 | 600 | 0.5 | 0.45 | 0.225[a] | Local device (T-shaped) |
| **$Bi_{45}Se_{55}$/Pt** [31] | 3-5 | 300 | 3700 | 0.35 | 3.2 | 1.12[a] | Local device (T-shaped) |
| **$Bi_xSe_{1-x}$/CoFeB** [19] | 4-40 | 300 | 12820 | | 18.62 | | Harmonic Hall (DC) |
| **$Bi_xSe_{1-x}$/YIG** [21] | 4-16 | 300 | | | | 0.11 | Spin pumping |
| **$Bi_xSe_{1-x}$/CoFeB** [20] | 2-16 | 300 | | | | 0.32 | Spin pumping |

[a] $\lambda_{IEE}\ (nm) = \theta_{SH}^{BiSe}\lambda_s^{BiSe}$. [b] Value taken from Ref. [31].

## IV. CONCLUSIONS

We successfully injected a pure spin current into highly resistive sputtered BiSe using non-local spin valves and performed spin absorption measurements from 10 K up to room temperature. A 3D FEM analysis of the absorption data allowed us to extract the spin diffusion length $\lambda_s^{BiSe}$ in this system for the first time. Spin-charge interconversion measurements were performed on the same device to extract the spin Hall angle, $\theta_{SH}^{BiSe}$. From these two experiments, we were able to reliably obtain the $\theta_{SH}^{BiSe}\lambda_s^{BiSe}$ product, a relevant figure of merit characterizing SCI in MESO devices. Despite the uncertainty regarding the resistivity of the Ti layer separating the Cu spin channel and the sputtered BiSe, the obtained values, generally lower than 1 nm, are robust. Although existing literature has reported a high SCI efficiency for BiSe and put this material forward as a promising candidate for MESO logic devices, our work shows otherwise. A more accurate characterization, relying on non-local devices that eliminate spurious effects, reveals that the SCI efficiency of sputtered BiSe is in fact too small to be used for MESO technology.

## ACKNOWLEDGMENTS

This work is supported by Intel Corporation through the Semiconductor Research Corporation under MSR-INTEL TASK Grant No. 2017-IN-2744 and the “FEINMAN” Intel Science Technology Center, and by the Spanish MICINN (Project No. PID2021-122511OB-I00 and Maria de Maeztu Units of Excellence Programme No. CEX2020-001038-M). W.Y.C. acknowledges postdoctoral fellowship support from “Juan de la Cierva—Formación” Programme by the Spanish MICINN (Grant No. FJC2018-038580-I). D.C.V acknowledges funding from the European Union’s Horizon 2020 research and innovation programme under the Marie Skłodowska-Curie Grant No. 892983-SPECTER.

# SUPPLEMENTAL MATERIAL

## Note S1

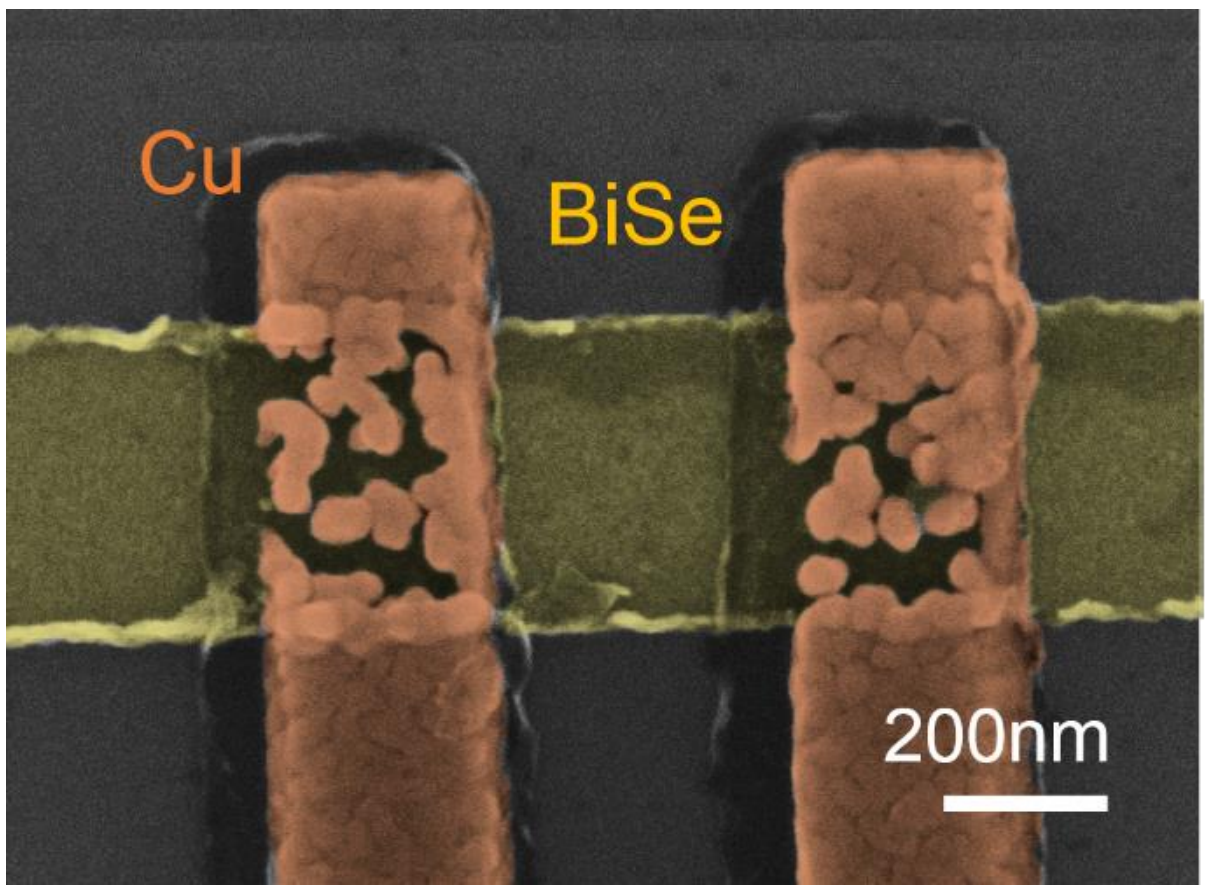

**Fig. S1 | a.** SEM image of two 100-nm-thick Cu wires on top of a 30-nm-thick BiSe wire. The image shows clearly how Cu does not grow as a homogeneous layer on top of BiSe, but as unconnected islands.

## Note S2

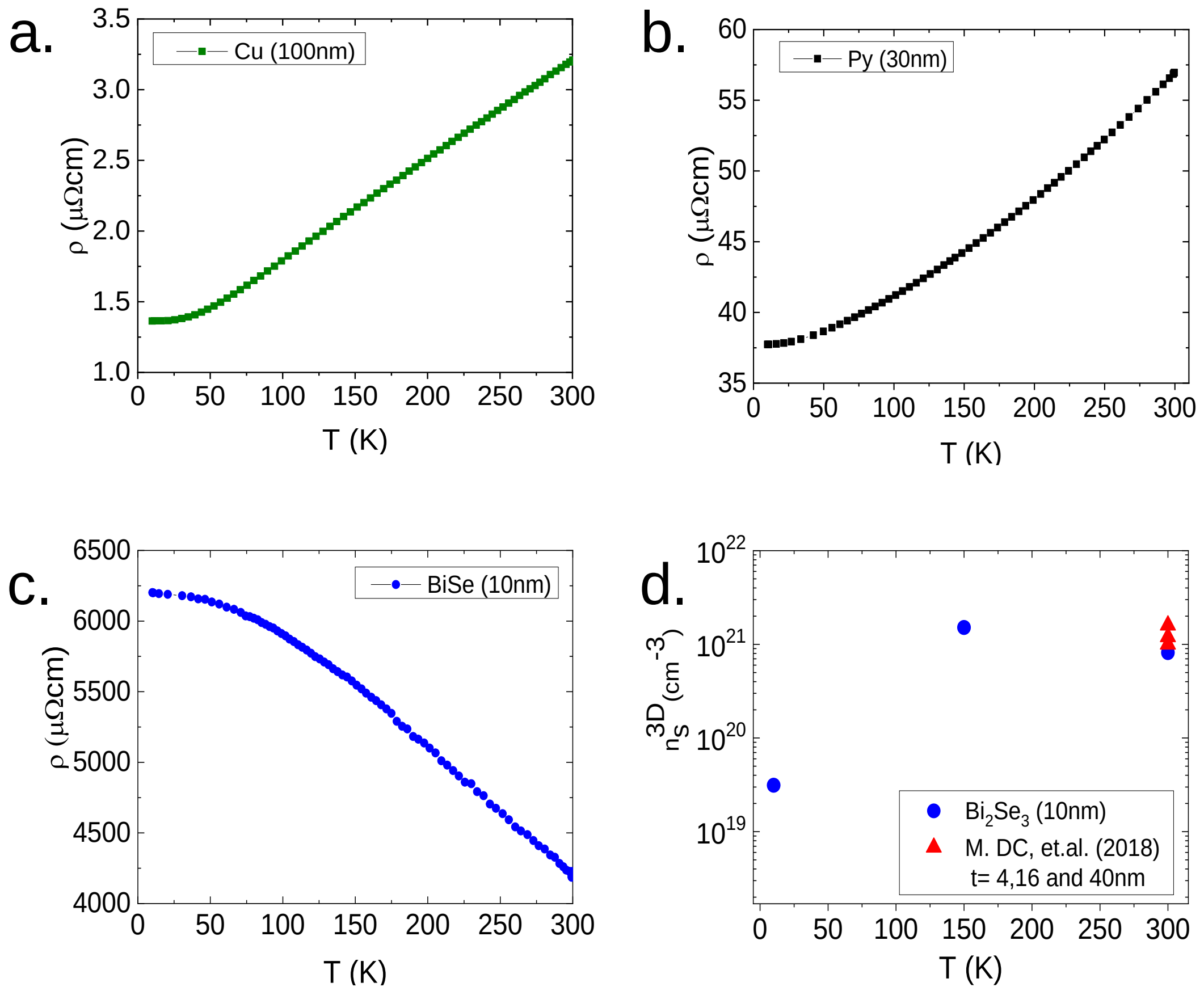

**Fig. S2 |** Measurements of resistivity as function of temperature for **a.** Cu (100 nm thick), **b.** Py (30 nm thick) and **c.** BiSe (10 nm). **d.** 3D carrier concentration, determined by Hall measurements, for BiSe (10 nm). Results from Ref. [19] are included for comparison.

**Note S3**

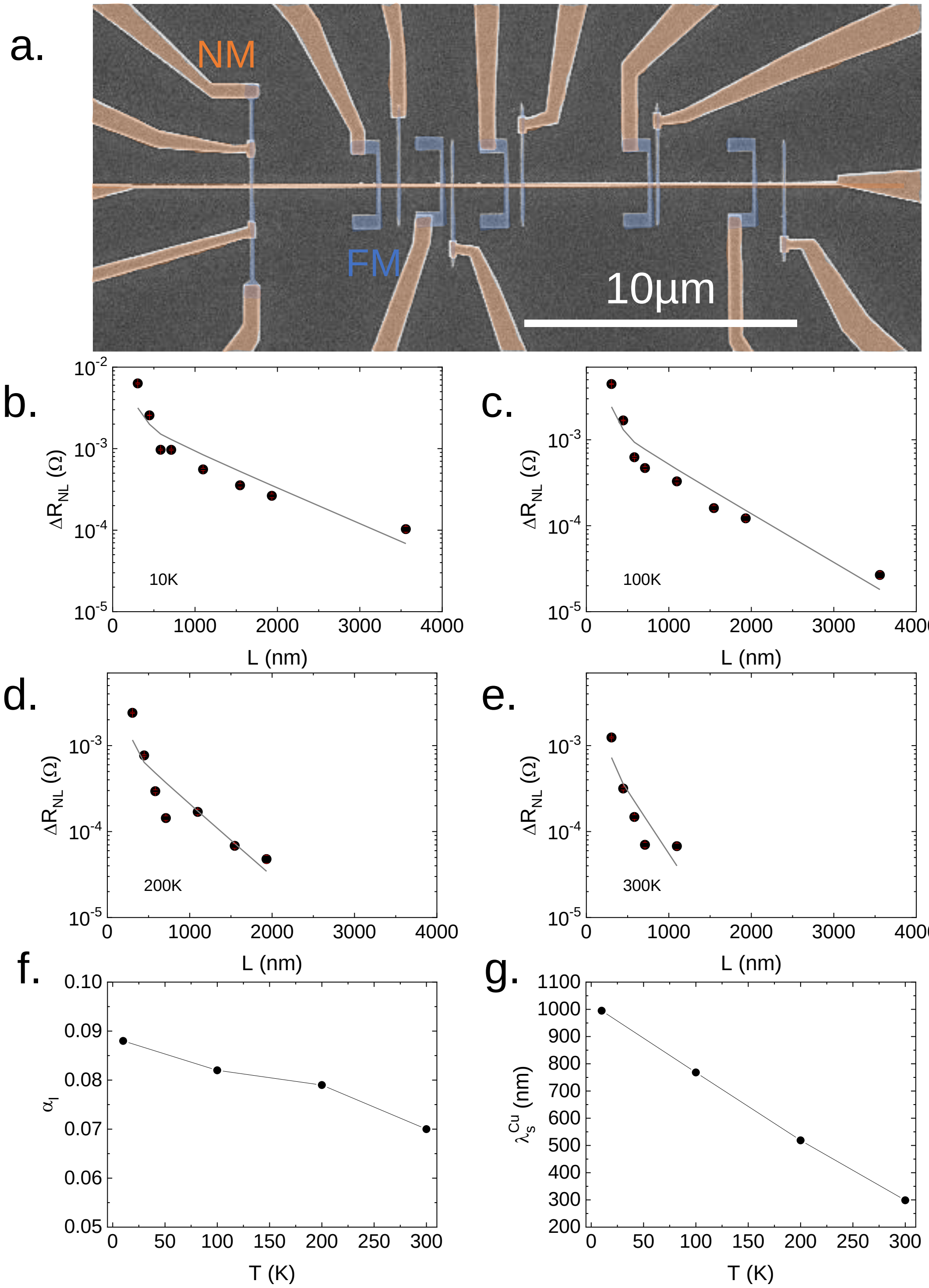

**Fig. S3 | a.** SEM image of LSVs with different distances between FM electrodes ($L$). FM (Py) electrodes are false colored in blue and the NM (Cu) spin transport channel in orange. A 2-nm-thick Ti layer is used between the Py and the Cu. The spin signal $\Delta R_{NL}$ as a function of $L$ at **b.** 10 K, **c.** 100 K, **d.** 200 K, and **e.** 300 K. Black circles are the experimental data (error bars are smaller than the symbol size), and solid grey lines the fit to Eq. 1. **f.** Interface spin polarization ($\alpha_I$) and **g.** spin diffusion length of Cu ($\lambda_s^{Cu}$) extracted from the fit at different temperatures.

In order to extract the spin diffusion length of Cu ($\lambda_s^{Cu}$) and the interface spin polarization ($\alpha_I$) of Py/Ti/Cu, on the same Si/$SiO_2$ chip as that containing the devices presented in the main text, but in a separate device (to guarantee the same experimentally conditions), we fabricated lateral spin valves (LSVs) with different Py electrode distances, as shown in Fig. S3a. The experimental signal ($\Delta R_{NL}$) decreases as both *L* and *T* increase. The $\Delta R_{NL}$ obtained are fitted following the 1D spin diffusion model for each temperature. Thus, the detected $\Delta R_{NL}$ is given by the general equation [54]:

$$\Delta R_{NL} = \frac{4R_S^{Cu}\left(\frac{\alpha_I}{1-\alpha_I{}^2}\frac{R_I}{R_S^{Cu}}+\frac{\alpha_{Py}}{1-\alpha_{Py}{}^2}\frac{R_S^{Py}}{R_S^{Cu}}\right)^2 e^{\frac{-L}{\lambda_S^{Cu}}}}{\left(1+\frac{1}{1-\alpha_I{}^2}\frac{R_I}{R_S^{Cu}}+\frac{1}{1-\alpha_{Py}{}^2}\frac{R_S^{Py}}{R_S^{Cu}}\right)^2 - e^{\frac{-2L}{\lambda_S^{Cu}}}} \quad (1)$$

where $R_s^{Cu(Py)}$ are the spin resistances and are defined as:

$$R_S^{Cu} = \rho_{Cu}\,\lambda_S^{Cu} / w_{Cu}t_{Cu} \quad (2)$$

$$R_S^{Py} = \rho_{Py}\,\lambda_S^{Py} / w_{Cu}w_{Py} \quad (3)$$

where $t_{Cu}$ is the thickness of the Cu channel, $\rho_{Cu(Py)}$ are the resistivities, $\lambda_s^{Cu(Py)}$ the spin diffusion lengths, and $w_{Cu(Py)}$ are the widths of the Cu channel and Py electrodes, respectively. $L$ is the distance between injector and detector electrodes. $\alpha_{Py}$ is the spin polarization of the Py and, together with $\lambda_s^{Py}$, these values are taken from Ref. [36]. $R_I$ is the interface resistance and is extracted as explained in Note S4.

## Note S4

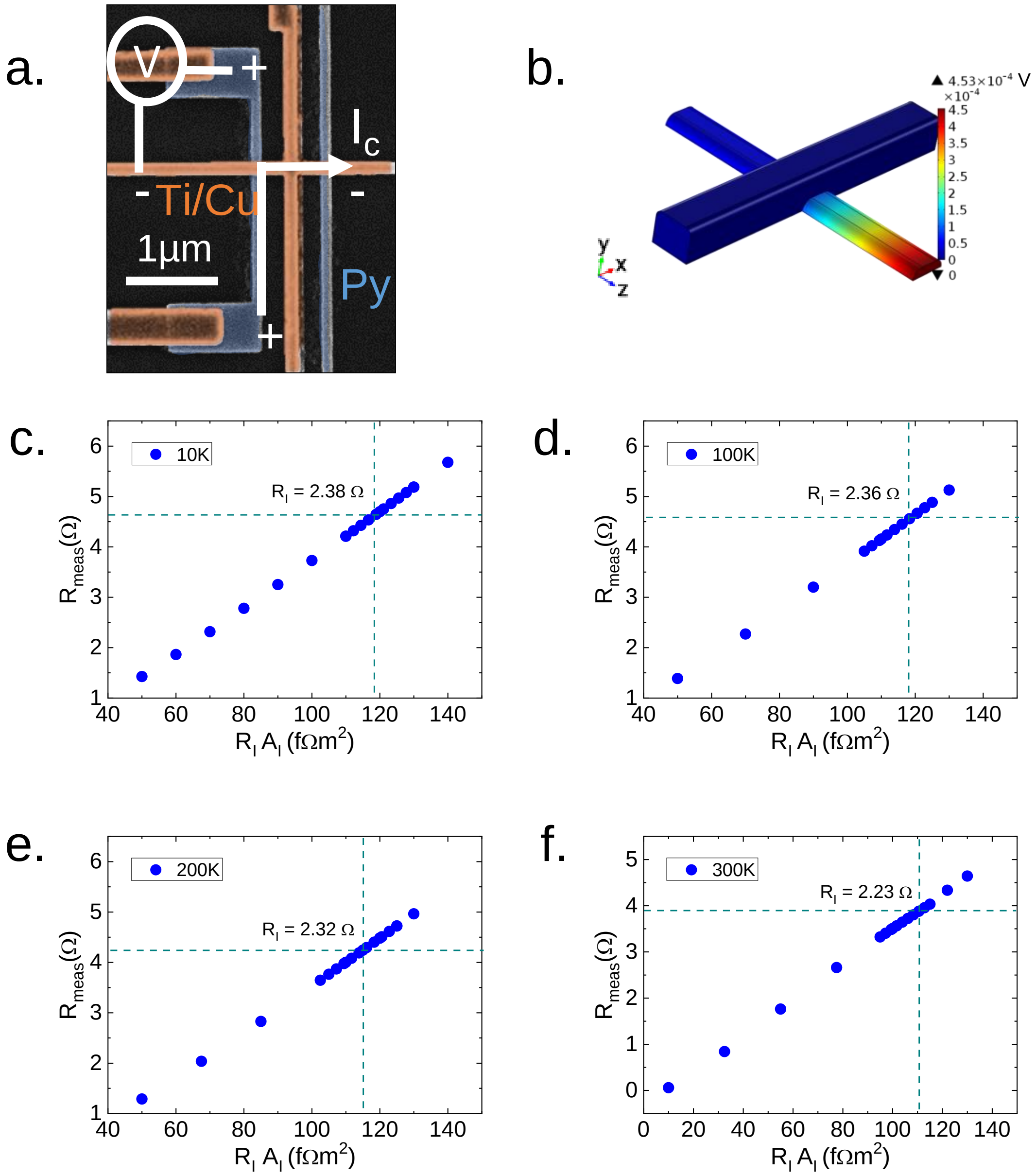

**Fig. S4 | a.** SEM image of the LSV devices presented in the main text showing the four-point electrical configuration used to extract the interface resistance of Py/Ti/Cu (in the injection or detection area, both are considered equal). **b**. 3D FEM simulation model presenting the geometry and the electrical potential driven by the applied current to extract the interface resistance ($R_I$). We extracted $R_I$ by adjusting this resistance in the interface area ($R_I A_I$, where $A_I = 5.0 \times 10^{-14}$ m$^2$) using the FEM calculation to reproduce the experimentally measured resistance $R_{meas}$. In this simulation, we considered Ti not as an extra layer but as an interface and used the resistivities of Cu and Py wires for each temperature: **c.** 10 K, **d.** 100 K, **e.** 200 K, and **f.** 300 K.

## Note S5

To extract the resistivity of the Ti layer on top of the BiSe, we fabricated a set of BiSe/Ti/Au stacks, varying the Ti thickness ($t_{Ti}$ = 1, 2, 3, 4 and 5 nm, the sketch is shown in the inset of Fig. S5), keeping the BiSe thickness constant ($t_{BiSe}$= 10 nm), and adding Au as a capping layer ($t_{Au}$= 3 nm). The stacks are patterned as Hall bars (width $w$ = 100 nm and length $L$ = 800 nm) to measure the total resistance ($R_{Tot}$), whose inverse is plotted as a function of $t_{Ti}$ in Fig. S5. By applying the parallel resistance model:

$$\frac{L}{w\,R_{Tot}} = \frac{t_{BiSe}}{\rho_{BiSe}} + \frac{t_{Ti}}{\rho_{Ti}} + \frac{t_{Au}}{\rho_{Au}} \tag{4}$$

where $\rho_{Ti,BiSe,Au}$ correspond to the resistivity of Ti, BiSe and Au, respectively, we can extract the resistivity of the Ti layer from the linear slope of the data in Fig. S5 (1 nm thick of Ti is too thin to allow for continuous growth, then is not considered for the fitting). Hall Bars of BiSe (10 nm)/Au (3nm) were made to measure the resistivity of the 3-nm-thick of Au ($\rho_{Au}$= 34.1 μΩcm).

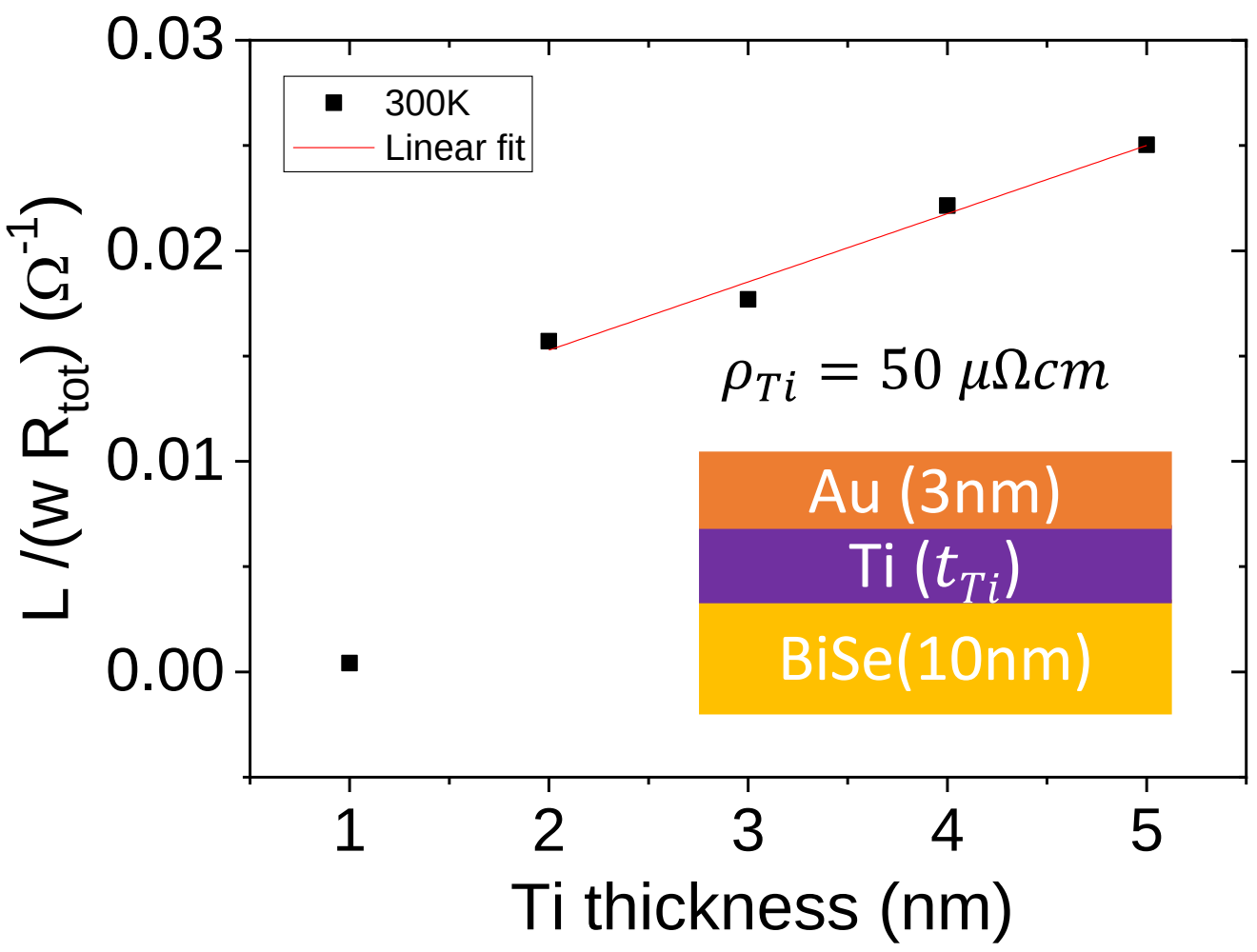

**Fig. S5 |** Inverse of the total resistance of the BiSe (10 nm)/Ti ($t_{Ti}$)/Au (3 nm) stack as a function of the Ti thickness. Inset: Sketch of the stack. Measurements performed at 300 K.

## Note S6

To extract the spin diffusion length of BiSe ($\lambda_s^{BiSe}$), we simulated the spin absorption measurements at 10 K, 100 K, 200 K, and 300 K by using a 3D finite element method (FEM) based on the two-current drift-diffusion model [55]. Figures S6a and S6b show the geometry of the simulated device (the actual device can be seen in Fig. 1a of the main text) and the mesh of the finite elements. The geometry construction and 3D-mesh were made using the free software GMSH [56] with the associated solver GETDP [57] for calculations, post- processing and together with a Phyton code to control data flow. Input parameters were taken from the resistivity characterization of each material shown in Note S2, $\lambda_s^{Cu}$ and $\alpha_I$ extracted as shown in Note S3, and interface resistance of the Py/Ti/Cu junction extracted as shown in Note S4. The resistivity of the 2-nm-thick Ti layer between Cu and BiSe wires was estimated from the control experiment described in Note S5. $\lambda_s^{Ti}$ was calculated assuming the Elliott-Yafet mechanism [58,59] for spin relaxation ($\rho_{Ti}\lambda_s^{Ti}$= cnt.) and taking the constant from Ref. [60] ($\lambda_s^{Ti}$= 13.3 nm for $\rho_{Ti}$=300 μΩcm). We simulated the spin absorption experiment by adjusting $\lambda_s^{BiSe}$ in the FEM calculation to reproduce the experimental spin signal in the absorption device ($\Delta R_{NL}^{abs}$, shown in figure 1c of the main text). The simulation is repeated for the experimental $\Delta R_{NL}^{abs}$ values at each temperature (shown in Fig. 1 of the main text). The output results are shown in Figs. S6c-f.

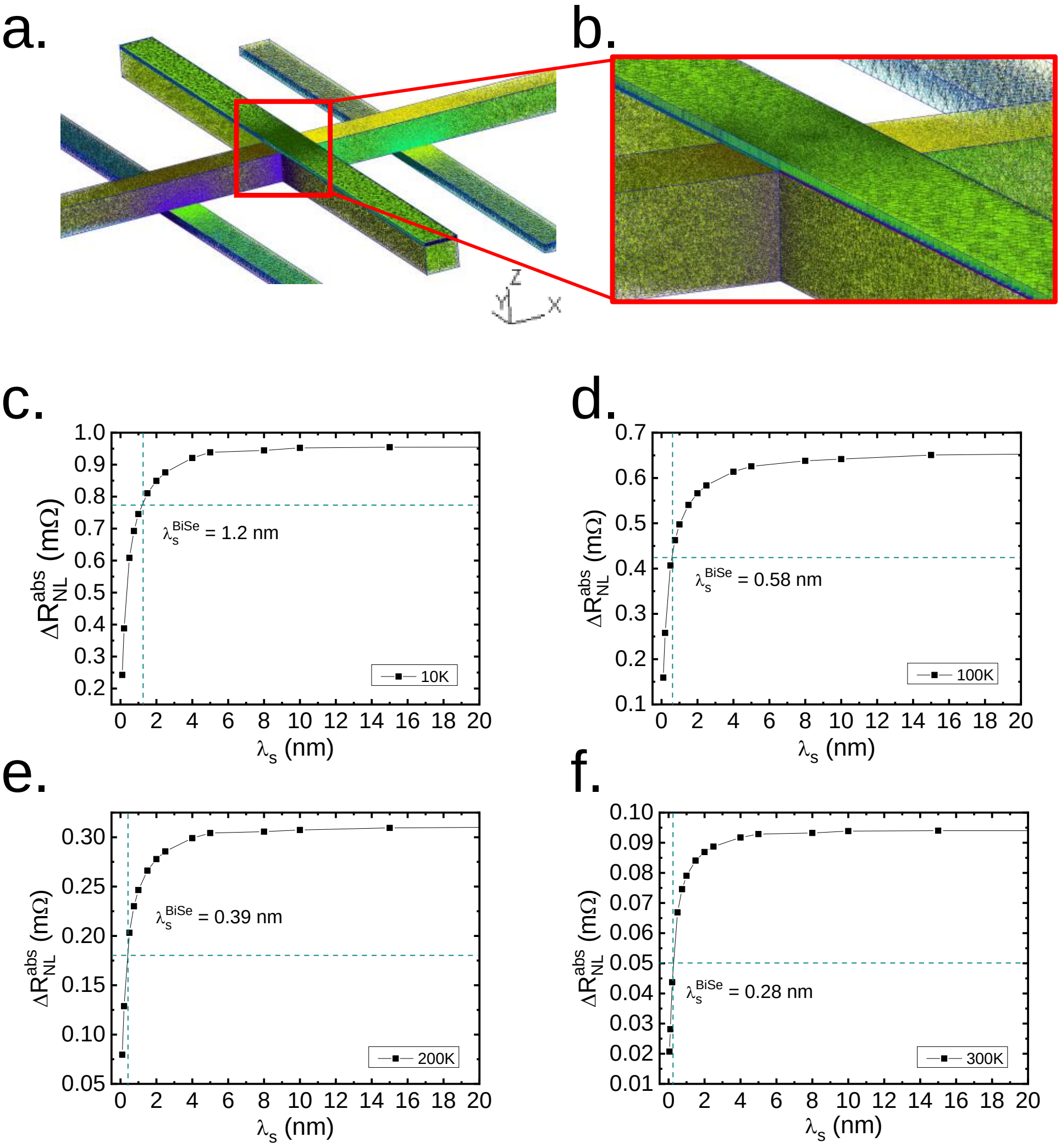

**Fig. S6 | a.** Geometry of the simulated device and the mesh of the finite elements and **b.** zoom on the 10-nm-thick BiSe wire with the 2-nm-thick Ti (purple) and 100-nm-thick Cu layers at the bottom. 3D FEM analysis output to extract the spin diffusion length of BiSe ($\lambda_s^{BiSe}$) considering a $\rho_{Ti}$ of 50 μΩcm at **c.** 10 K, **d.** 100 K, **e.** 200 K, and **f.** 300 K.

## Note S7

To extract the spin Hall angle of BiSe ($\theta_{SH}^{BiSe}$), a 3D FEM simulation is performed in a similar manner as the one in Note S6 but for the spin-to-charge conversion measurements at 10 K, 100 K, 200 K, and 300 K. Since we use the same device as for spin absorption, the geometry is the same, but the electrical contacts are changed (see Fig. 2 of the main text). Then, by adjusting the effective spin Hall angle, $\theta_{SH}^{eff}$, in the simulation to reproduce the experimental value ($\Delta R_{(I)SHE}$, shown in Fig. 2e of the main text) and using the same input parameters as in Note S6 and the obtained value of $\lambda_s^{BiSe}$ (presented in Note S6 and shown in Fig. 1e of the main text), we obtained $\theta_{SH}^{BiSe}$ for each temperature.

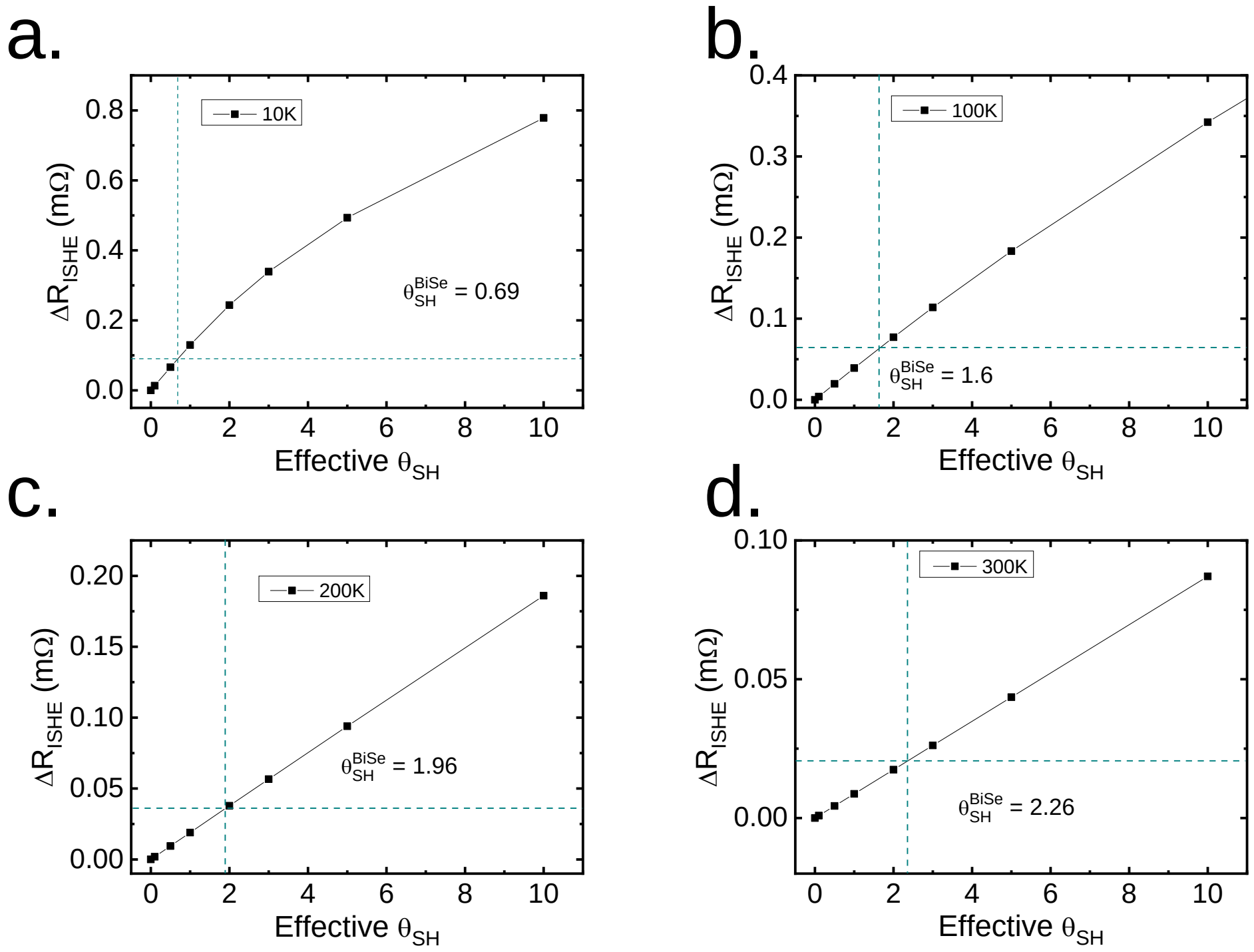

**Fig. S7 |** 3D FEM analysis output to extract the spin Hall angle of BiSe ($\theta_{SH}^{BiSe}$) considering a $\rho_{Ti}$ of 50 μΩcm at **a.** 10 K, **b.** 100 K, **c.** 200 K, and **d.** 300 K.

## Note S8

We performed a control spin-charge interconversion experiment in the reference device (i.e., without the BiSe wire under the Ti/Cu cross). In this reference device, no spin-charge interconversion is expected. The experiment is performed under the same conditions as in the devices with the BiSe wire. Figure S8 shows the SHE resistance as a function of the magnetic field. No SHE signal is observed above the noise level of $\sim 5\ \mu\Omega$ , indicating that the presence of SHE is negligible.

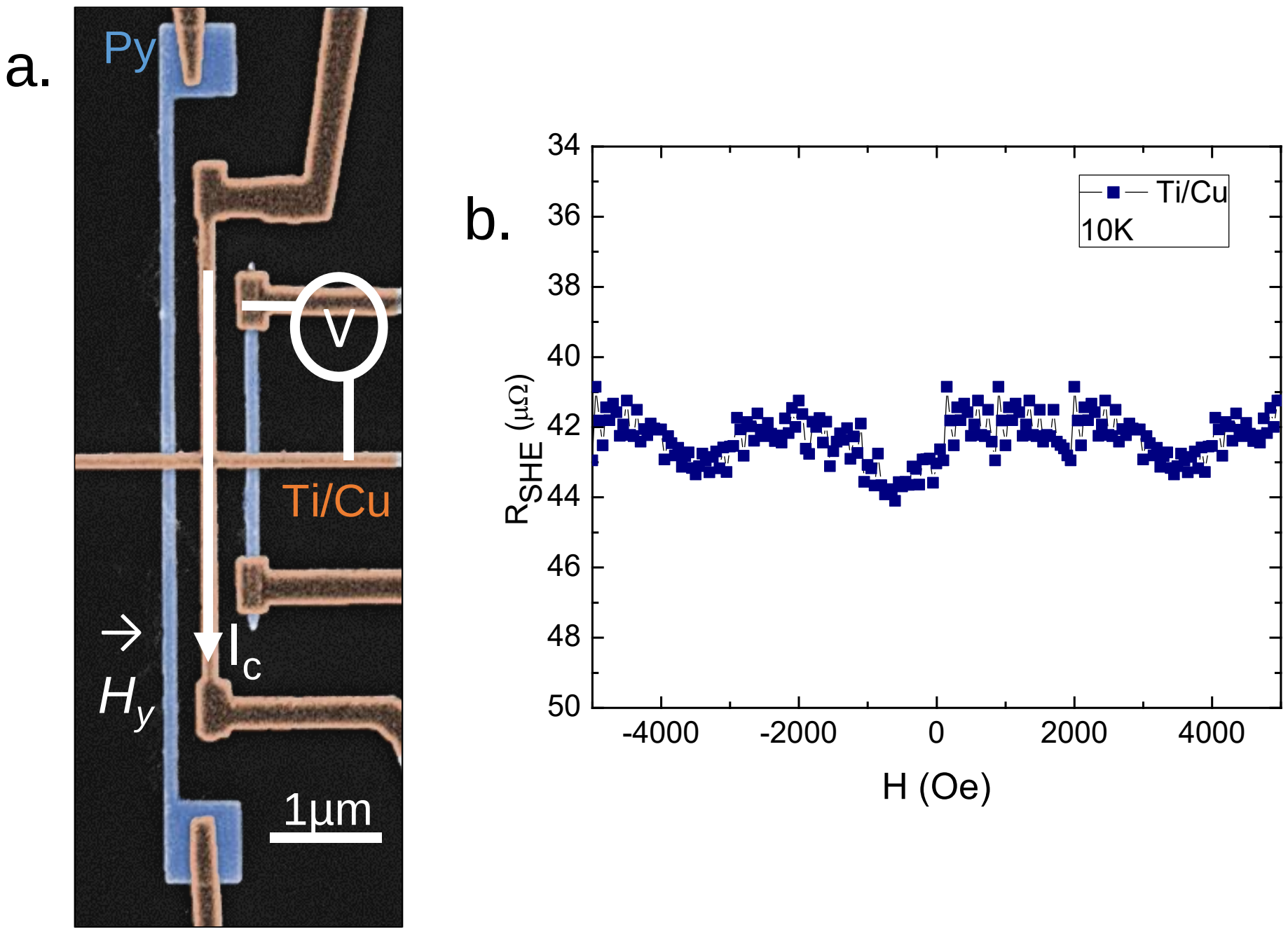

**Fig. S8 | a.** Top-view SEM image of the reference LSV with the electrical configuration for the SHE measurement. The control experiment consists in injecting a charge current directly in the cross of Ti (2 nm)/ Cu (100 nm), while reading the voltage between the Py electrode and the Cu channel and applying the magnetic field in the hard axis of the Py electrode. **b.** SHE resistance in Ti/Cu as a function of the external magnetic field at 10 K, measured with the configuration shown in panel a. No SHE signal is observed above the noise level.

## Note S9

We performed a cross section cut to study the interfaces by TEM and EDX in the very same device where we performed the spin absorption and the spin Hall measurements. By performing an EDX elemental analysis, it is possible to clearly observe the Ti layer (2 nm) in between the BiSe wire (10 nm) and the Cu channel (100 nm), as well as the presence of oxygen in the Ti layer. The detailed results are shown in Fig. S9.

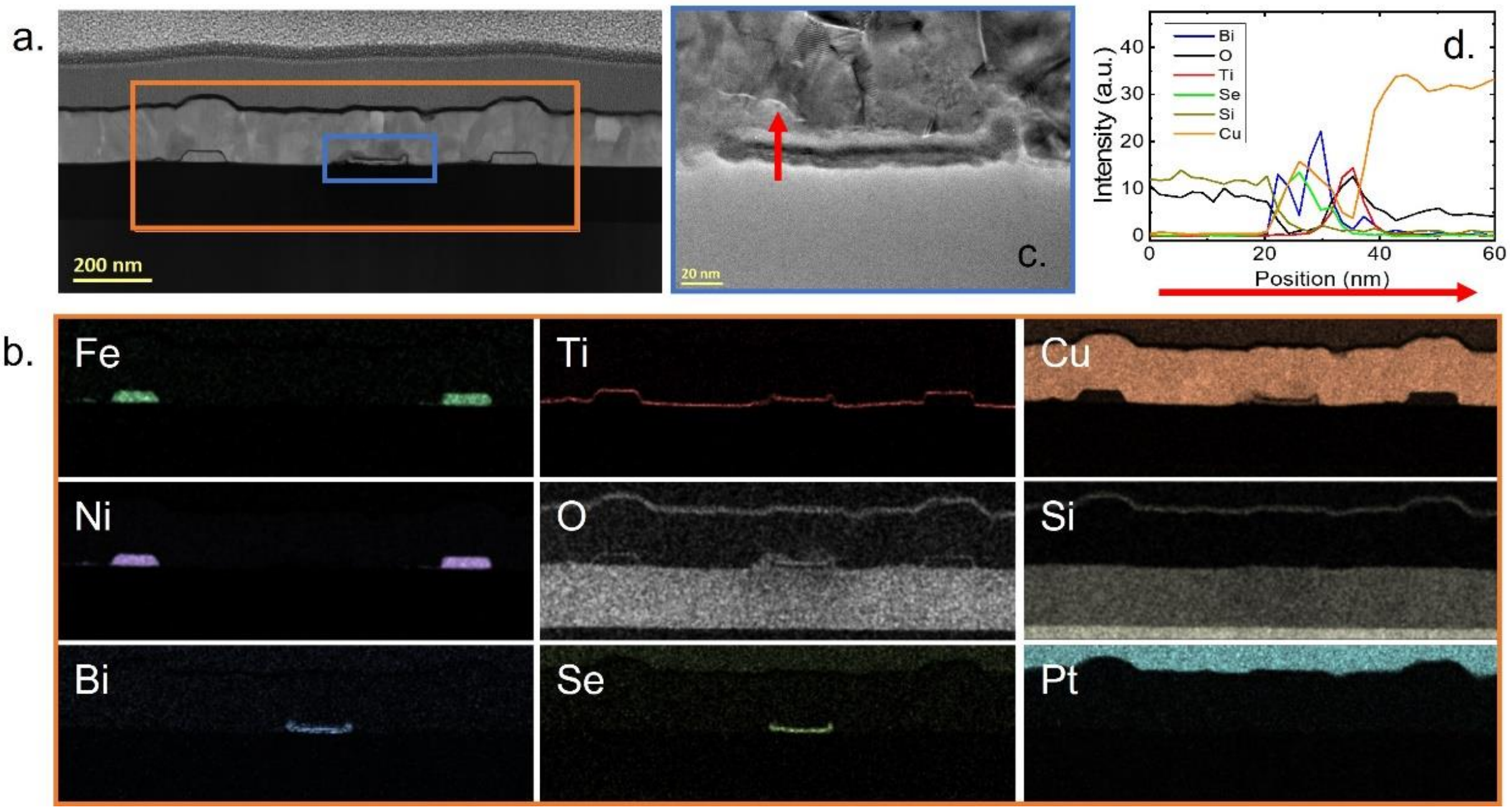

**Fig. S9 | a.** Cross-sectional TEM image of the LSV used in this work. **b.** EDX analysis for each relevant element present in the sample, performed in the area marked with an orange box in panel a. **c.** High resolution TEM image of the BiSe wire region (blue box in panel a). **d.** Elemental profile through the BiSe wire. The red arrow represents the scanning direction (also labeled in panel c), from the substrate ($Si/SiO_2$) to the top of the sample (Ti/Cu).

## Note S10

We repeated the 3D FEM simulation to extract $\lambda_s^{BiSe}$ at 10 K, 100 K, 200 K, and 300 K by considering the Ti layer might be oxidized. As we presented in Fig. 3 of the main text and Fig. S9, the Ti layer contains oxygen. However, extracting the resistivity of this layer is not possible in the geometry of the device, because the BiSe wire is fully shunted with the Ti/Cu channel. The resistivity ranges used in the simulation are chosen as follows: The lower limit is based on the measured value of a control experiment described in Note S5 (50 $\mu\Omega$cm). The upper limit is based on the extracted resistivity assuming the interface resistance at the Py/Ti/Cu junction (1000 $\mu\Omega$cm). Nevertheless, it is important to highlight that each material grows differently on top of different materials. When the Ti layer is simulated to be too resistive ($\rho_{Ti}$=1500 μΩcm), $\lambda_s^{BiSe}$ tends to infinity, meaning that fewer spins can reach the BiSe layer, to the point where the properties of the BiSe layer are no longer significant in the 3D model. The output results for each temperature are shown in Figs. S10-S13.

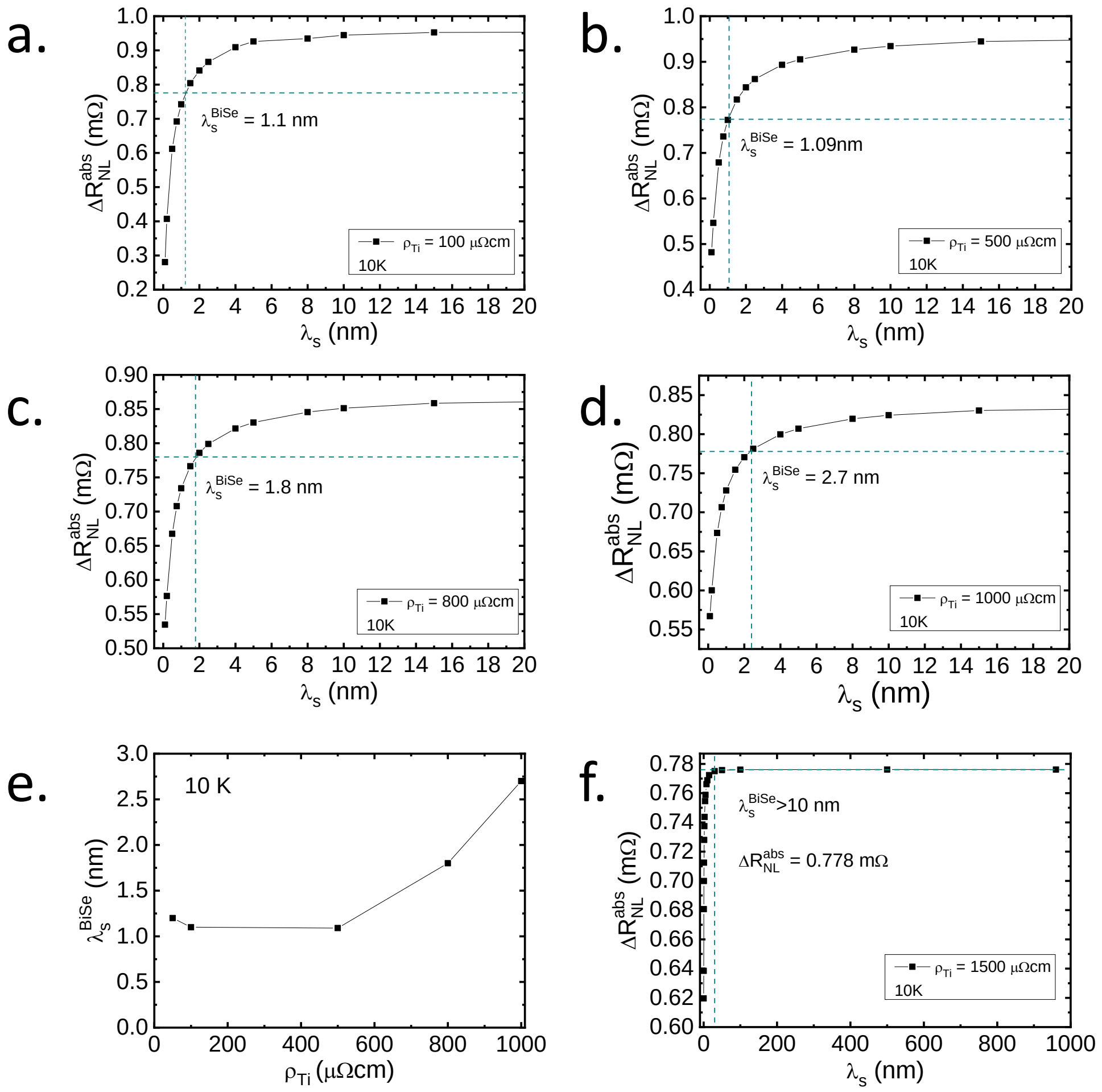

**Fig. S10 |** 3D FEM analysis output to extract $\lambda_s^{BiSe}$ at 10 K considering a $\rho_{Ti}$ of **a.** 100 μΩcm, **b.** 500 μΩcm, **c.** 800 μΩcm, and **d.** 1000 μΩcm. **e.** $\lambda_s^{BiSe}$ as a function of the Ti resistivity at 10 K, obtained from the analysis in panels a-d and Fig. S6c. **f.** 3D FEM analysis output to extract $\lambda_s^{BiSe}$ at 10 K assuming a high Ti resistivity of 1500 μΩcm.

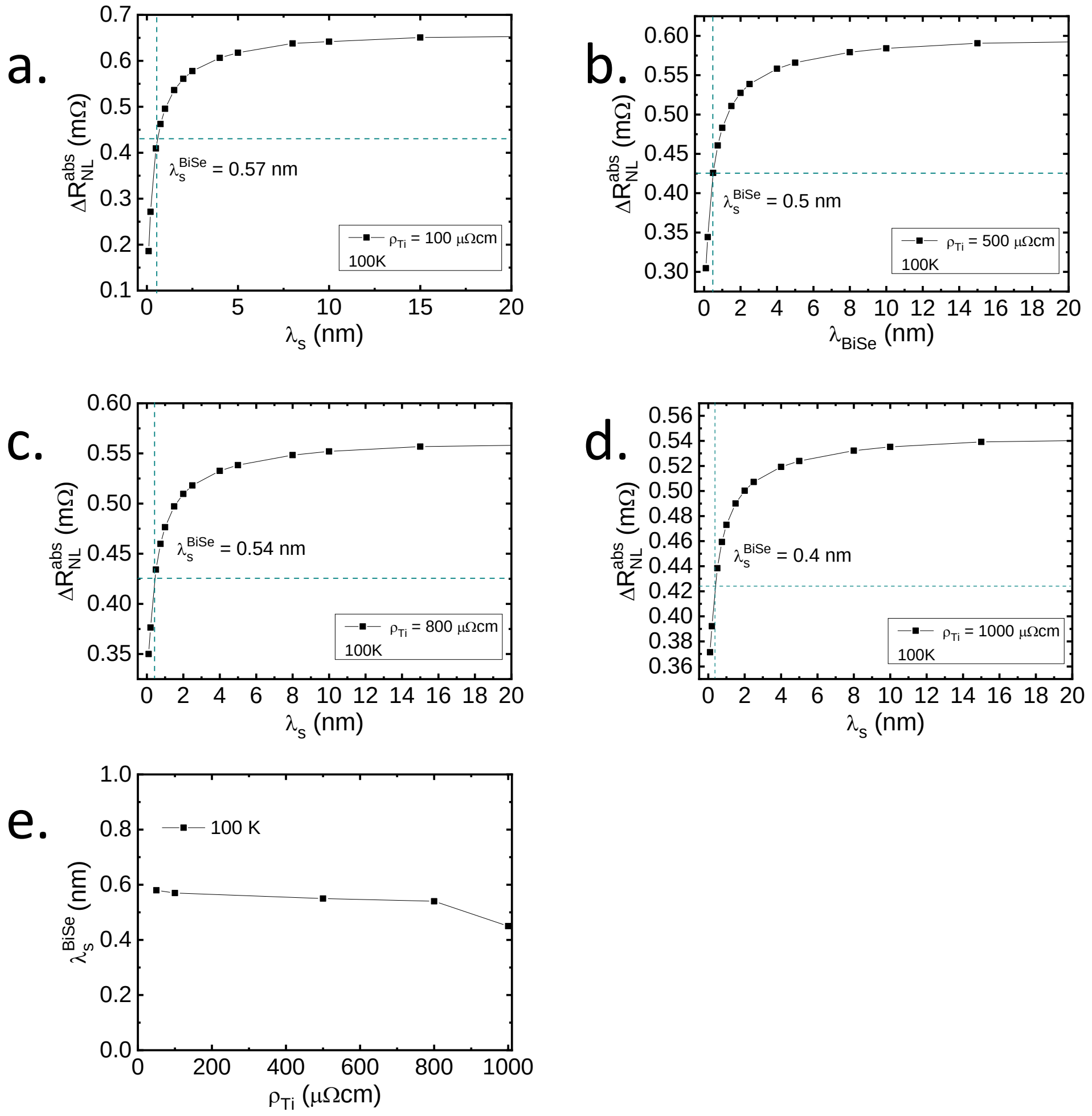

**Fig. S11 |** 3D FEM analysis output to extract $\lambda_s^{BiSe}$ at 100 K considering a $\rho_{Ti}$ of **a.** 100 μΩcm, **b.** 500 μΩcm, **c.** 800 μΩcm, and **d.** 1000 μΩcm. **e.** $\lambda_s^{BiSe}$ as a function of the Ti resistivity at 100 K, obtained from the analysis in panels a-d and Fig. S6d.

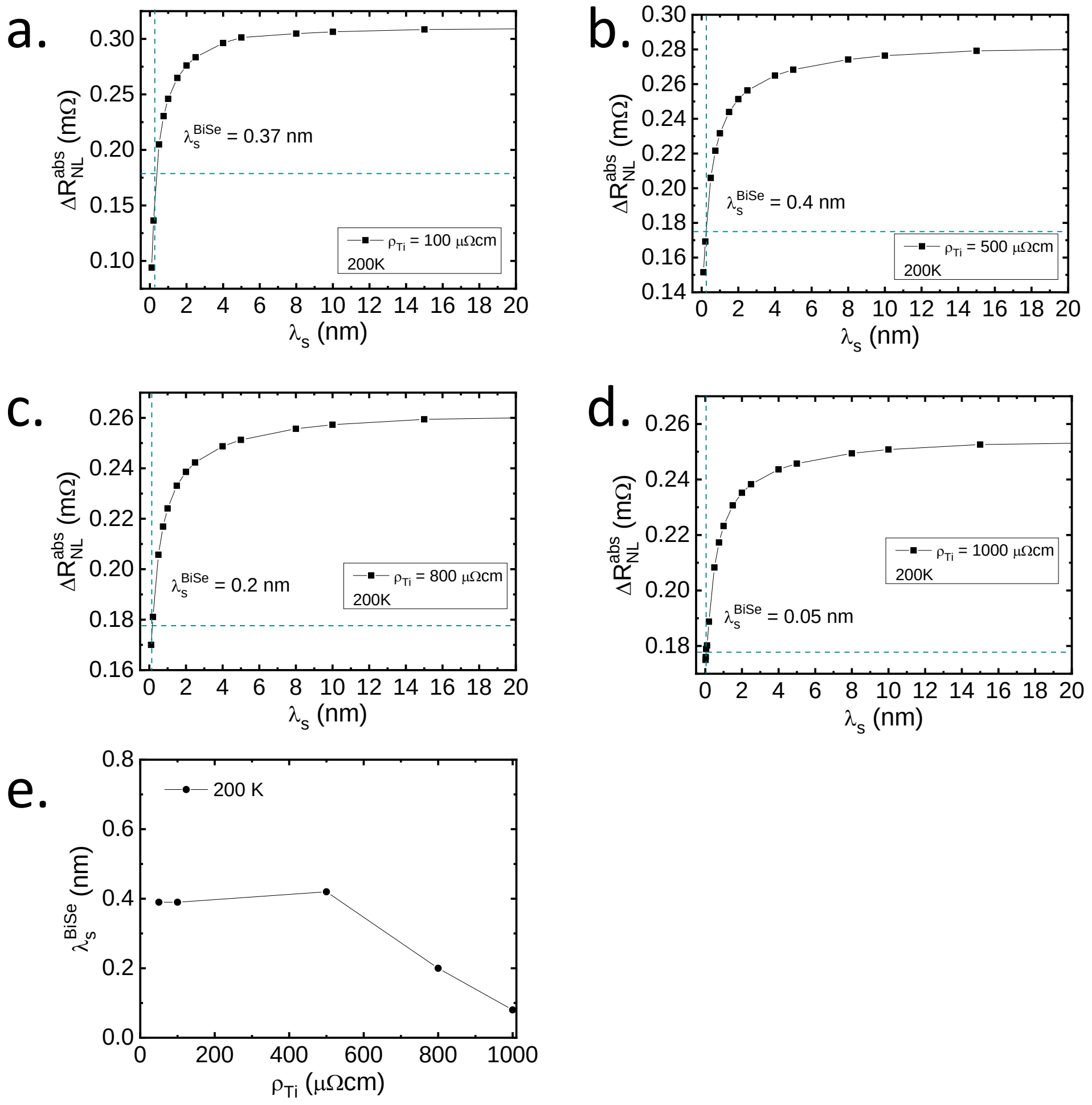

**Fig. S12 |** 3D FEM analysis output to extract $\lambda_s^{BiSe}$ at 200 K considering a $\rho_{Ti}$ of **a.** 100 μΩcm, **b.** 500 μΩcm, **c.** 800 μΩcm, and **d.** 1000 μΩcm. **e.** $\lambda_s^{BiSe}$ as a function of the Ti resistivity at 200 K, obtained from the analysis in panels a-d and Fig. S6e.

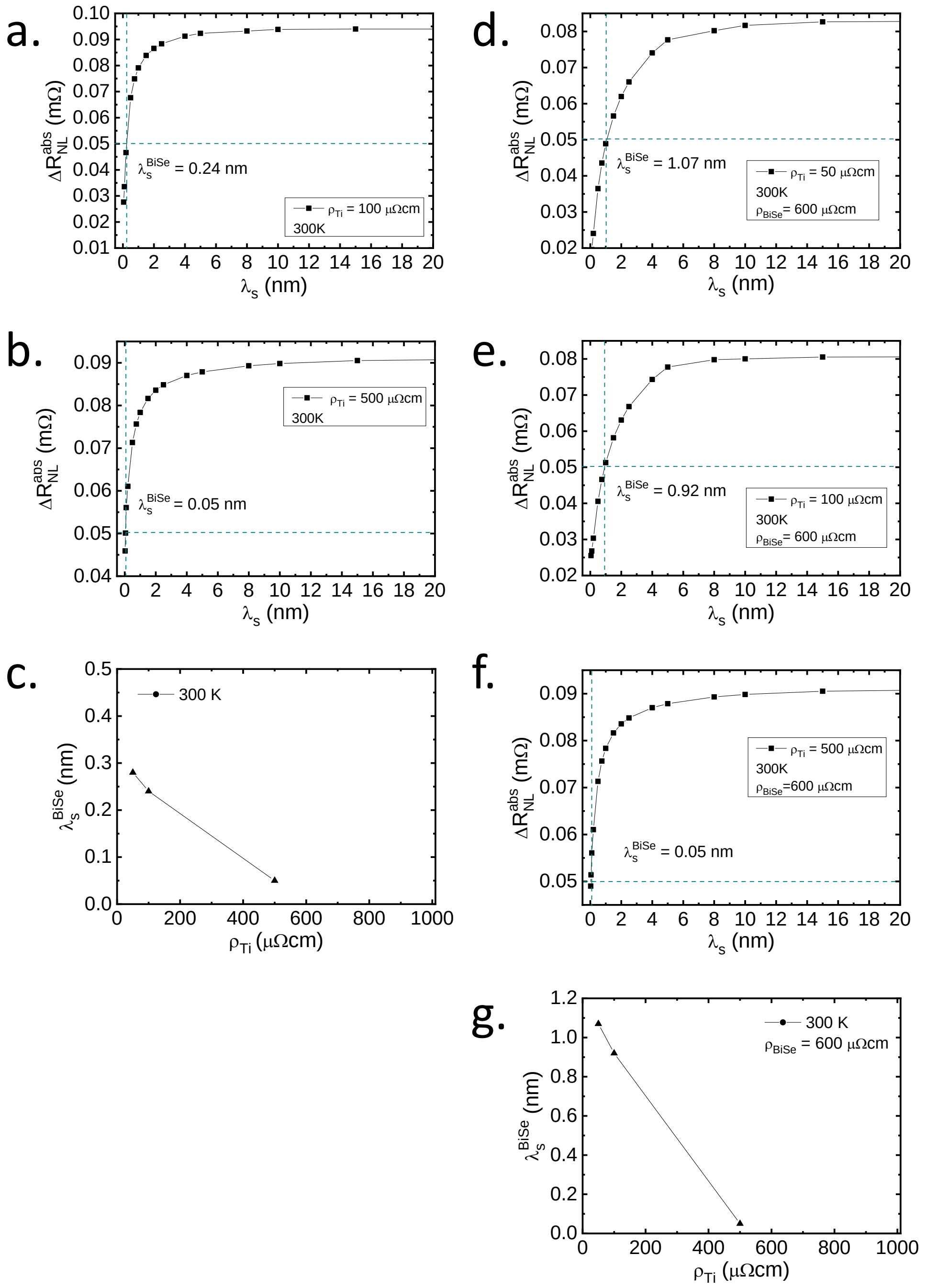

**Fig. S13 |** 3D FEM analysis output to extract $\lambda_s^{BiSe}$ at 300 K considering a $\rho_{BiSe}$ of 4100 μΩcm and a $\rho_{Ti}$ of **a.** 100 μΩcm and **b.** 500 μΩcm. **c.** $\lambda_s^{BiSe}$ as a function of the Ti resistivity at 300 K, obtained from the analysis in panels a-b and Fig. S6f. 3D FEM analysis output to extract $\lambda_s^{BiSe}$ at 300 K considering a lower BiSe resistivity ($\rho_{BiSe}$= 600 μΩcm, taken from Ref. [31]) and a $\rho_{Ti}$ of **d.** 50 μΩcm, **e.** 100 μΩcm, and **f.** 500 μΩcm. **g.** $\lambda_s^{BiSe}$ considering a $\rho_{BiSe}$ of 600 μΩcm as a function of the Ti resistivity at 300 K, obtained from the analysis in panels d-f.

## Note S11

We repeated the 3D FEM simulation to extract $\theta_{SH}^{BiSe}$ at 10 K, 100 K, 200 K and 300 K by considering different Ti resistivities, similar to that in Note S10, but for the spin-to-charge measurements at 10 K, 100 K, 200 K, and 300 K. Then, by adjusting the effective spin Hall angle, $\theta_{SH}^{eff}$, in the simulation to reproduce the experimental value ($\Delta R_{(I)SHE}$ shown in Fig. 2e of the main text) and using the same input parameters as in Note S6 and the obtained value of $\lambda_s^{BiSe}$ (presented in Note S10), we obtained $\theta_{SH}^{BiSe}$ for each temperature. The output results for each temperature are shown in Figs. S14-S17.

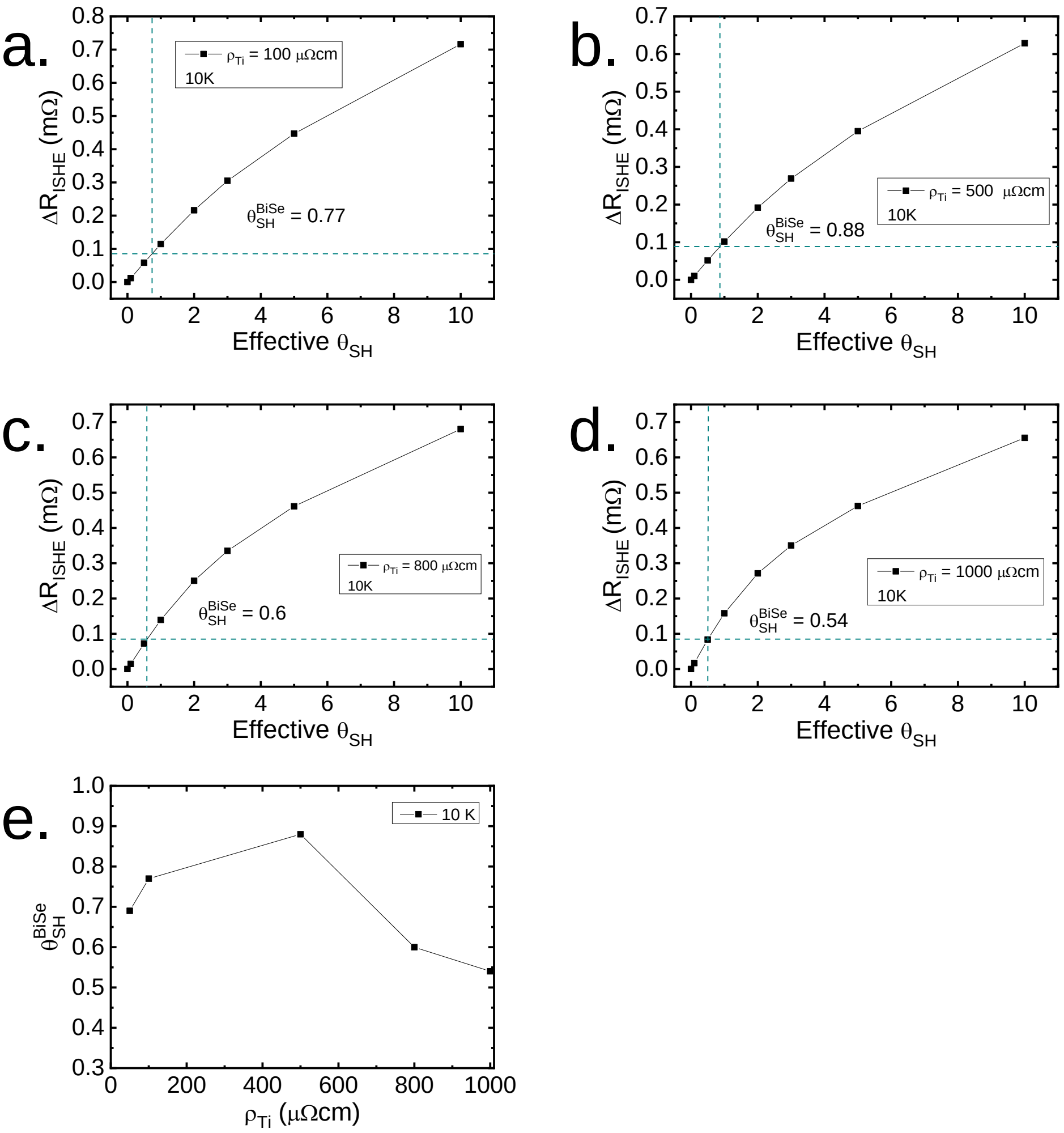

**Fig. S14 |** 3D FEM analysis output to extract $\theta_{SH}^{BiSe}$ at 10 K considering a $\rho_{Ti}$ of **a.** 100 μΩcm, **b.** 500 μΩcm, **c.** 800 μΩcm, and **d.** 1000 μΩcm. **e.** $\theta_{SH}^{BiSe}$ as a function of the Ti resistivity at 10 K, obtained from the analysis in panels a-d and Fig. S7a.

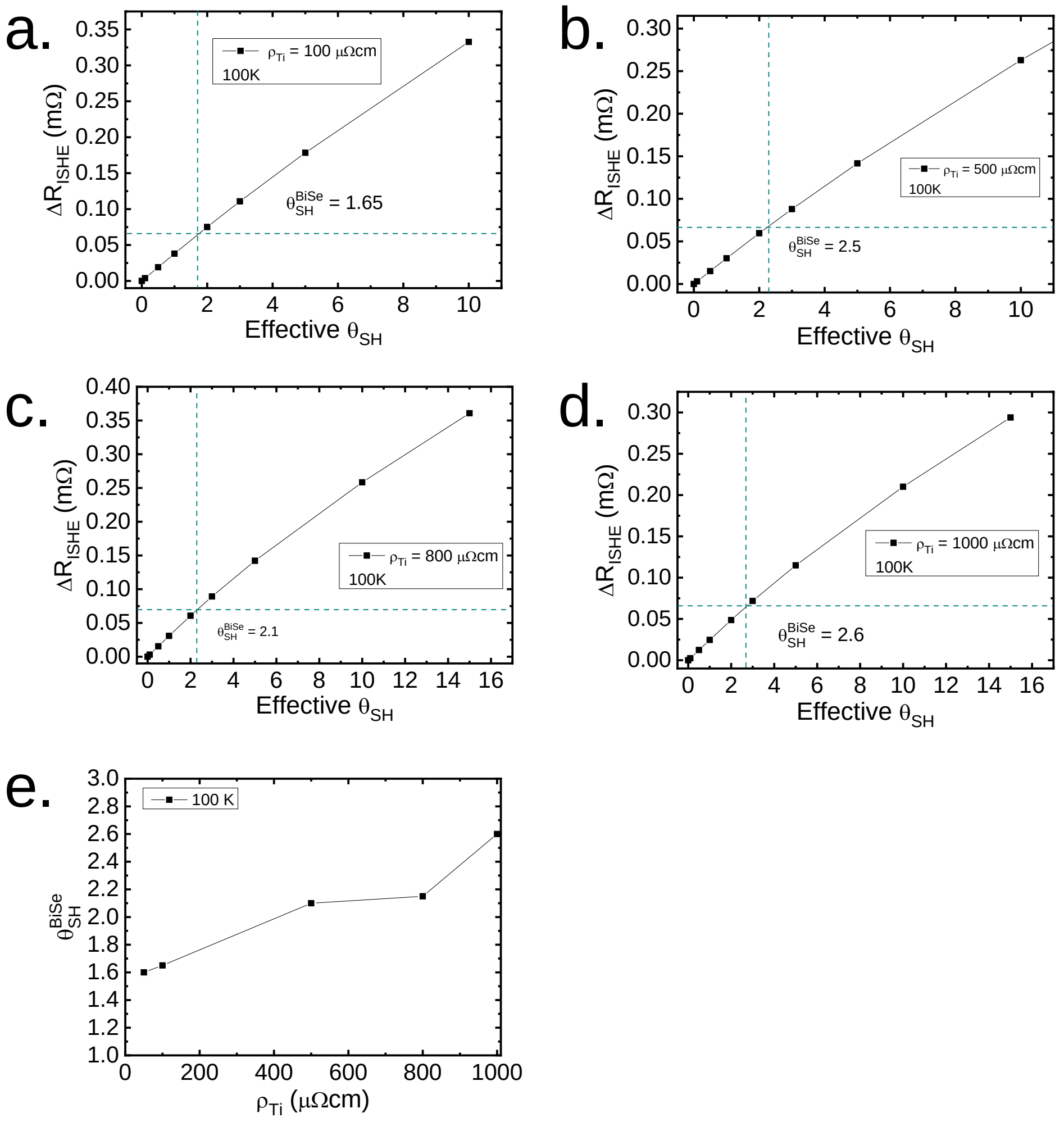

**Fig. S15 |** 3D FEM analysis output to extract $\theta_{SH}^{BiSe}$ at 100 K considering a $\rho_{Ti}$ of **a.** 100 μΩcm, **b.** 500 μΩcm, **c.** 800 μΩcm, and **d.** 1000 μΩcm. **e.** $\theta_{SH}^{BiSe}$ as a function of the Ti resistivity at 100 K, obtained from the analysis in panels a-d and Fig. S7b.

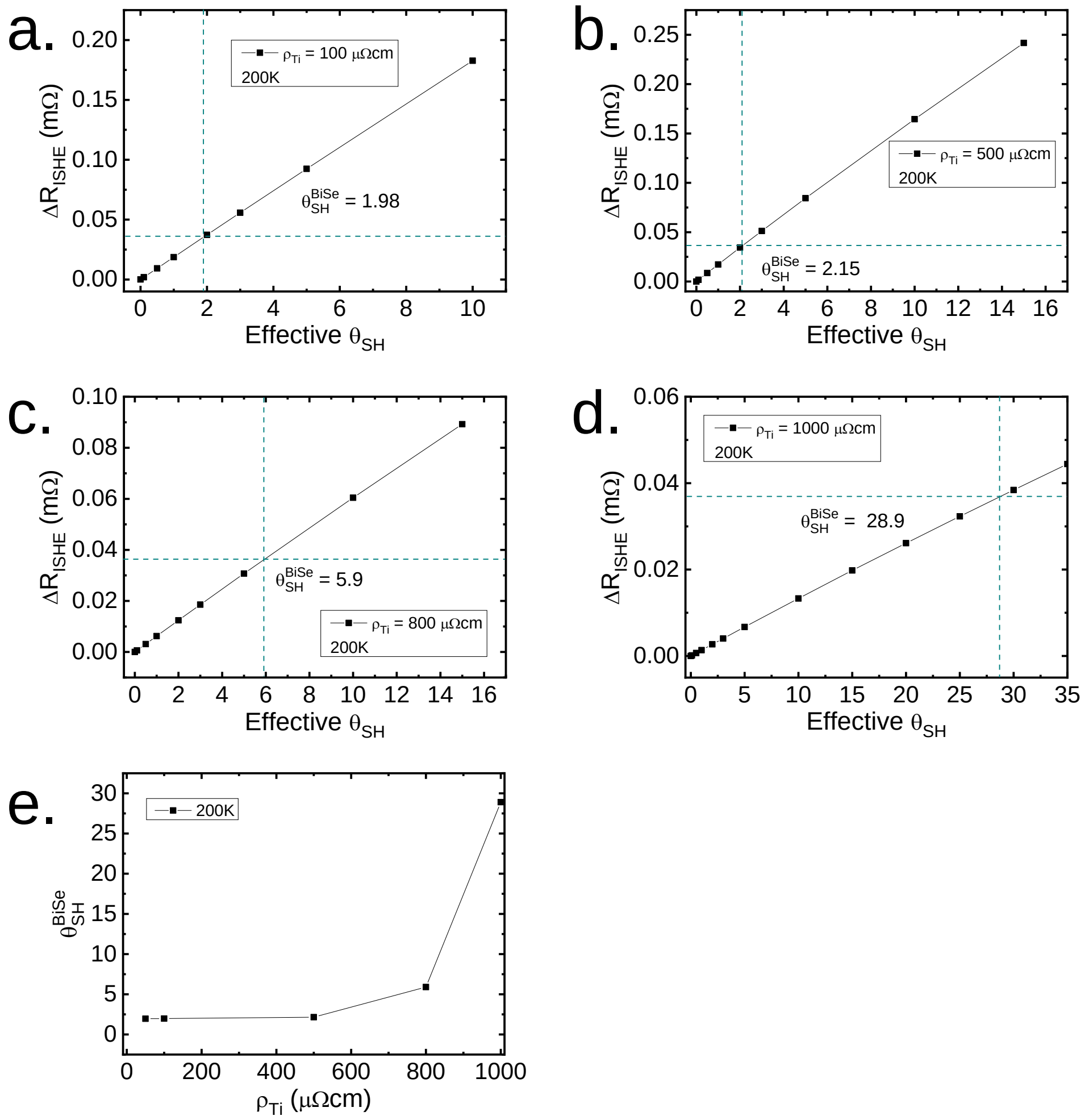

**Fig. S16 |** 3D FEM analysis output to extract $\theta_{SH}^{BiSe}$ at 200 K considering a $\rho_{Ti}$ of **a.** 100 μΩcm, **b.** 500 μΩcm, **c.** 800 μΩcm, and **d.** 1000 μΩcm. **e.** $\theta_{SH}^{BiSe}$ as a function of the Ti resistivity at 200 K, obtained from the analysis in panels a-d and Fig. S7c.

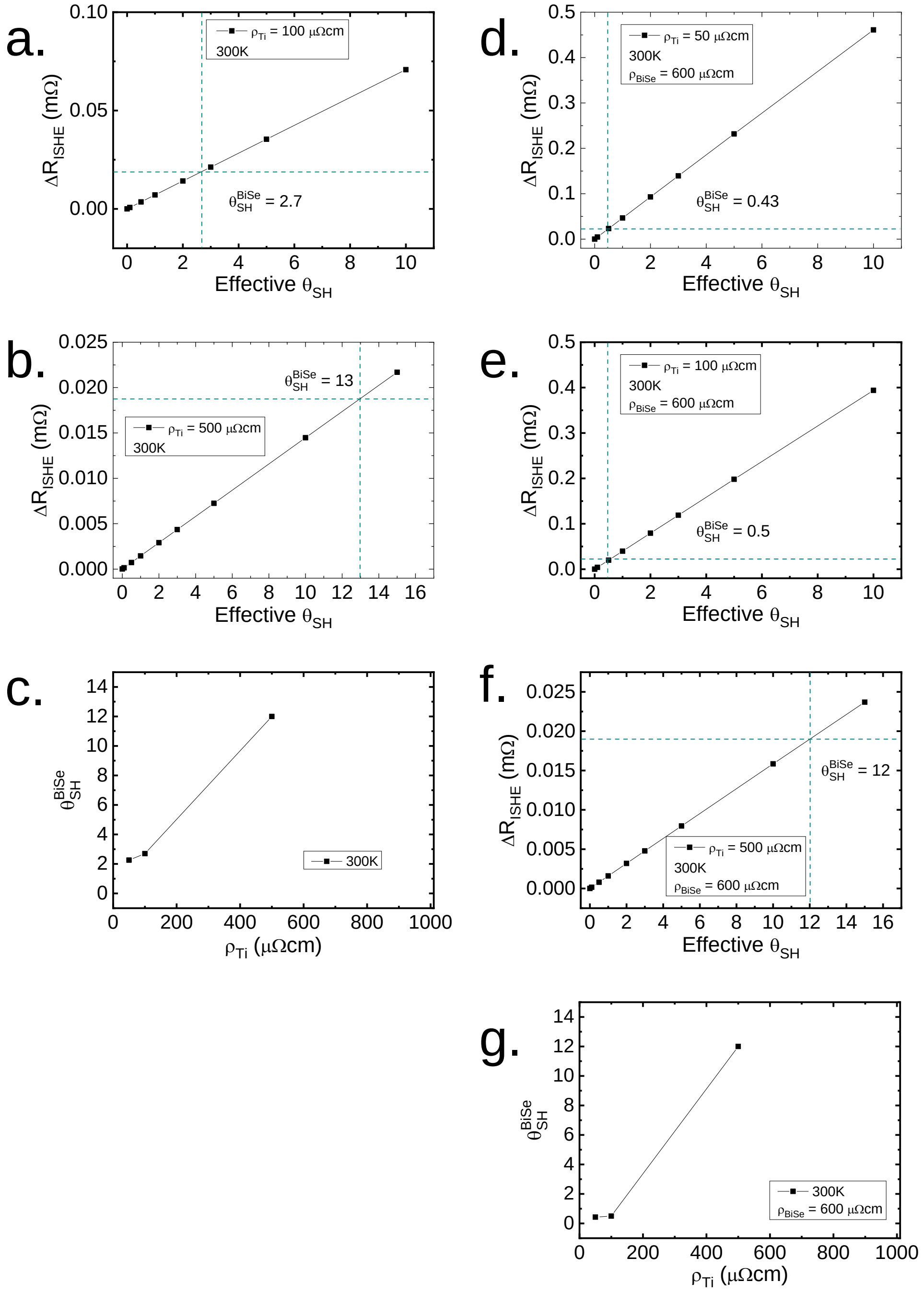

**Fig. S17 |** 3D FEM analysis output to extract $\theta_{SH}^{BiSe}$ at 300 K considering a $\rho_{BiSe}$ of 4100 μΩcm and a $\rho_{Ti}$ of **a.** 100 μΩcm and **b.** 500 μΩcm. **c.** $\theta_{SH}^{BiSe}$ as a function of the Ti resistivity at 300 K, obtained from the analysis in panels a-b and Fig. S7d. 3D FEM analysis output to extract $\theta_{SH}^{BiSe}$ at 300 K considering a lower BiSe resistivity ($\rho_{BiSe}$= 600 μΩcm, taken from Ref. [31]) and a $\rho_{Ti}$ **d.** 50 μΩcm, **e.** 100 μΩcm, and **f.** 500 μΩcm. **g.** $\theta_{SH}^{BiSe}$ considering a $\rho_{BiSe}$ of 600 μΩcm as a function of the Ti resistivity at 300 K, obtained from the analysis in panels d-f.